\documentclass{aa}
\usepackage{graphicx}
\usepackage[varg]{txfonts}
\usepackage{natbib}
\bibpunct{(}{)}{;}{a}{}{,}
\usepackage{color}
\usepackage{mathrsfs}
\definecolor{comm}{rgb}{0,0.5,0}

\definecolor{old}{rgb}{0.6,0.4,0.4}

\definecolor{new}{rgb}{0.5,0.2,0.5}

\definecolor{done}{rgb}{1,0.5,0}

\definecolor{todo}{rgb}{0.8,0,0}

\definecolor{idea}{rgb}{0.3,0.5,1}

\newcommand{\cd}{d$^{-1}$}

\allowdisplaybreaks
\begin{document} 

\title{Tidally excited oscillations in MACHO\,80.7443.1718: Changing amplitudes and frequencies, high-frequency tidally excited mode, and a decrease in the orbital period}
  \author{P. A. Ko{\l}aczek-Szyma\'nski\inst{1},
  A. Pigulski\inst{1},
  M. Wrona\inst{2},
  M. Ratajczak\inst{2}
  \and
  A. Udalski\inst{2}
}
  \institute{Astronomical Institute, University of Wroc\l aw, Kopernika 11, 51-622 Wroc\l aw, Poland\\
  \email{kolaczek@astro.uni.wroc.pl}
  \and
  Astronomical Observatory, University of Warsaw, Al. Ujazdowskie 4, 00-478 Warszawa, Poland\\
}
  \date{Received 7 September 2021; accepted 12 November 2021}
\abstract
{Eccentric ellipsoidal variables (also known as heartbeat stars) is a class of eccentric binaries in which proximity effects, and tidal distortion due to time-dependent tidal potential in particular, lead to measurable photometric variability close to the periastron passage. Varying tidal potential may also give rise to tidally excited oscillations (TEOs), which are forced eigenmodes with frequencies close to the integer multiples of the orbital frequency. TEOs may play an important role in the dynamical evolution of massive eccentric systems.}
{Our study is aimed at detecting TEOs and characterising the long-term behaviour of their amplitudes and frequencies in the extreme-amplitude heartbeat star MACHO\,80.7443.1718, consisting of a blue supergiant and a late O-type massive dwarf.} 
{We used two seasons of Transiting Exoplanet Survey Satellite (TESS) observations of the target to obtain new 30-min cadence photometry by means of the difference image analysis of TESS full-frame images. In order to extend the analysis to longer timescales, we supplemented the TESS data with 30-year long ground-based photometry of the target. Both TESS and ground-based photometry are carefully analysed by means of Fourier techniques in order to detect TEOs, examine the long-term stability of their amplitudes and frequencies, and characterise other types of variability in the system.}
{We confirm the detection of the known $n=23$, 25, and 41 TEOs and announce the detection of two new TEOs, with $n=24$ and 230, in the photometry of MACHO\,80.7443.1718. Amplitudes of all TEOs were found to vary on a timescale of years or months. For $n=25$, the TEO amplitude and frequency changes are related, which may indicate that the main cause of the amplitude drop in this TEO in TESS observations is the change in its frequency and increase in its detuning parameter. The light curve of the $n=230$ TEO is strongly non-sinusoidal. Its high frequency may indicate that the oscillation is a strange mode. Stochastic variability observed in the target fits the behaviour observed in massive stars well and independently confirms that the primary is an evolved star. We also find that the orbital period of the system  decreases at a rate of about 11\,s\,(yr)$^{-1}$. This can be explained by several phenomena: a significant mass loss, mass transfer between components, tidal dissipation, and the presence of a tertiary in the system. All of these phenomena may contribute to the observed changes.}
{The discovery of variable amplitudes and frequencies of TEOs prompts for similar studies in other eccentric elliptical variables with TEOs. Long-term photometric monitoring of these targets is also desirable. The results we obtained pose a challenge for theory. In particular, it needs to be explained why $n=230$ TEO is excited. In a general context, studies on the long-term behaviour of TEOs may help to explain the role of TEOs in the dynamical evolution of massive eccentric systems.}
\keywords{binaries: close -- stars: early-type -- stars: massive -- stars: oscillations -- stars: individual: MACHO\,80.7443.1718}
\titlerunning{Tidally excited oscillations in MACHO\,80.7443.1718}
\authorrunning{Ko{\l}aczek-Szyma\'nski et al.}
\maketitle
%

\section{Introduction}\label{sect:introduction}

Massive main-sequence stars frequently reside in binary and multiple systems \citep{2013ARA&A..51..269D}. Since they are also young, systems they belong to usually did not have enough time to circularise their orbits. In effect, many massive stars are found in highly eccentric systems. This gives rise to many interesting phenomena related to the tidal interaction between the components including mass transfer during periastron passages. Proximity effects, in particular tidal distortion and mutual irradiation, lead to the occurrence of significant photometric variability. Although observed earlier in many eccentric (and eclipsing) binaries \citep[see, e.g.][]{1986A&A...160..310G,2000ApJ...544..409G,2007IBVS.5782....1V}, photometric variability close to the periastron passage did not attract much attention until some extreme cases were discovered with the Kepler satellite photometry \citep{2012ApJ...753...86T}. In a highly eccentric system, variability due to proximity effects is confined to a narrow range in orbital phase close to the periastron and sometimes resembles an electrocardiogram. Thus, the term `heartbeat stars' (HBSs) has been coined for stars showing this effect. In principle, however, it is better to call these stars eccentric ellipsoidal variables (EEVs) to also include stars with lower eccentricities in which a variable signal due to proximity effects is smaller and does not change so rapidly. Throughout this paper, we therefore refer to these stars as EEVs, although we use the term `a heartbeat' for the short part of a light curve of an EEV close to a periastron passage in which photometric changes are the fastest.

The history of the theoretical investigation of tidal effects in eccentric binaries and especially tidally excited oscillations (hereafter TEOs) is long \citep[see, e.g.][]{1975A&A....41..329Z,1999A&A...341..842W,2017MNRAS.472.1538F}, but numerous discoveries of EEVs with TEOs in Kepler and then in Transiting Exoplanet Survey Satellite (TESS) data triggered a lot of interest in this topic. Many papers deal with searching for circumstances favourable for the occurrence of TEOs and predicting their amplitudes \citep{2012MNRAS.420.3126F,2017MNRAS.472.1538F}. Some address the role of TEOs in the dissipation of the orbital energy and dynamic evolution of a binary, especially when resonance locking occurs \citep{2017MNRAS.472L..25F,2017MNRAS.472.1538F,2021AJ....161..263Z}. A possibility of transferring energy from pulsations to orbit which may increase the non-synchronicity of the components or eccentricity of the system, the so-called inverse tides, was also found \citep{2020MNRAS.497.4363L,2021MNRAS.501..483F}. A review of both the theory and observations of TEOs in massive binaries has been recently presented by \cite{2021FrASS...8...67G}.

Detailed observational studies of a large sample of EEVs and their TEOs are still lacking, however. One of the interesting problems that needs to be addressed is the long-term behaviour of TEOs, their vanishing and (re)appearance, changes in amplitudes and phases, the related timescales, etc. This can be particularly interesting in the context of high mass-loss rates that can be enhanced by mass transfer in eccentric massive binaries. With the present study, we partly try to fill this gap by analysing all available photometry of an extreme case, that is to say the EEV with the largest known amplitude of the heartbeat, MACHO\,70.7443.1718. Because of this peculiarity, throughout this paper we use the abbreviation `ExtEEV' for this extreme EEV. The target is introduced below; we also explain why this star is particularly well suited for this kind of study.

The target star, MACHO\,70.7443.1718 ($V\sim$\,13.3\,mag), was discovered as a variable within the MAssive Compact Halo Object (MACHO; Sect.\,\ref{sect:macho}) survey of the Large Magellanic Cloud (LMC) and classified as an eclipsing binary with a period of about 32.83\,d \citep{2001yCat.2247....0M}. However, it was \cite{2019MNRAS.489.4705J} (hereafter Jaya19) who, in using All Sky Automated Survey for SuperNovae (ASAS-SN; Sect.\,\ref{sect:asassn}) ground-based data and  TESS space photometry from the first two sectors, identified the star as a very massive eccentric binary with an extremely large amplitude of the heartbeat ($\sim$0.4\,mag). In addition, TEOs were found in this star.

In the follow-up spectroscopic and photometric study, \cite{2021MNRAS.506.4083J}, hereafter Jaya21, confirmed that the object belongs to the LMC and classified the primary as a B-type supergiant, B0\,Iae, which is in good agreement with \cite{1994AJ....108.1256G}, who provided a similar classification, B0.5\,Ib-II. Unfortunately, Jaya21 did not find lines of the secondary in their spectra. Nevertheless, from the estimates for the luminosity and effective temperature and with the use of evolutionary models, they estimated the primary's mass for about 35\,M$_\odot$. A combination of the mass function derived for this single-lined spectroscopic binary, an inclination obtained from modelling the heartbeat ($i\approx 44\degr$), and the primary's mass allowed them to also estimate the secondary's mass for about 16\,M$_\odot$. Assuming coevality of the components, they conclude that the secondary is likely an O9.5\,V star. These authors also found evidence for a circumstellar disk and argue that the primary is a B[e] star. Using TESS data from 18 sectors available at the time of their analysis, they found two TEOs, one already known from the study of Jaya19 with $n=25$\footnote{Throughout the paper, we refer to TEOs using their $n$ numbers, which are defined via their frequencies $f_{\rm TEO} \approx nf_{\rm orb}$, where $f_{\rm orb}$ is the orbital frequency.} and another with $n=41$. The eccentricity of the system is relatively large, $e\approx 0.51$. The estimated age of the system, which is equal to about 6\,Myr, is consistent with the age of the parent OB association LH\,58 \citep{1994AJ....108.1256G}. All of these pieces of evidence led Jaya21 to conclude that the primary component of the ExtEEV has already left the main sequence and started its journey across the Hertzsprung gap.

For several reasons, the ExtEEV is a perfect target for studies aimed at testing the theory of the interaction between tidal effects and pulsations in massive stars. First of all, it is located in the TESS continuous viewing zone, which means that the extremely long record of TESS observations is available for this system. Next, it shows the strongest known heartbeat feature, which itself is worth studying and cannot be properly modelled with the present modelling codes (Jaya19, Jaya21). It also shows one of the strongest known TEOs, with a semi-amplitude of about 1 per cent or 10 parts-per-thousand (hereafter ppt) of the mean flux. Last but not least, the star is located in the LMC, which is a target of many ground-based photometric surveys. This allowed Jaya19 to discover it as a HBS and for us to study temporal behaviour of its strongest TEO (Sect.\,\ref{sect:ground-based}).

In the present paper, we first use TESS full-frame images (FFIs) to extract new photometry for the target star, then perform time-series analysis of these data, and discuss all types of detected variability (Sect.\,\ref{sect:tess-var}). Subsequently, we focus on the detected TEOs, in particular on the temporal behaviour of their amplitudes and phases (Sect.\,\ref{sect:changing-a-and-f}). Since we found that amplitudes and phases of TEOs vary, we included ground-based photometry to our analysis (Sect.\,\ref{sect:ground-based}) to check how the strongest TEO behaves on the timescale of about 30 years. The results are discussed in Sect.\,\ref{sect:discussion} and we conclude in the last section.


\section{Variability of the ExtEEV in TESS data}\label{sect:tess-var}
\subsection{TESS photometry}
TESS is a space-borne observatory collecting photometric data in a wide passband centred at $\sim$700\,nm and mainly dedicated to studies of exoplanets. With four separate cameras covering approximately $24\degr\,\times\,96\degr$ in the sky and pixel (hereafter shortened to `pix') scale of 21\,\arcsec, TESS delivered 2-min cadence light curves of selected objects and FFIs in 30-min intervals during the first two years of operation. In the ongoing extended time of operation, the FFI cadence has been shortened to 10 minutes and a new 20-s cadence has been added. The observations are performed in sectors, which are axially distributed with respect to the ecliptic, with the longest viewing zones centred around the ecliptic poles. Each sector is observed for 27 days, but because sectors partially overlap, an object can be observed for a longer time, depending on the angular distance from the ecliptic. The ExtEEV is located within the southern TESS continuous viewing zone, but it does not have 2-min cadence observations. Its photometry can only be obtained by using 30-min cadence FFIs obtained during the first year of the primary TESS mission in sectors 1\,--\,7, 9\,--\,10, and 12\,--\,13 and 10-min cadence FFIs obtained during the extended phase of the mission in sectors 27 and 29\,--\,36. Throughout our paper, we refer to the former group of observations as `TESS year 1' and to the latter one as `TESS year 3' data.
\begin{figure}
   \centering
   \includegraphics[width=0.75\hsize]{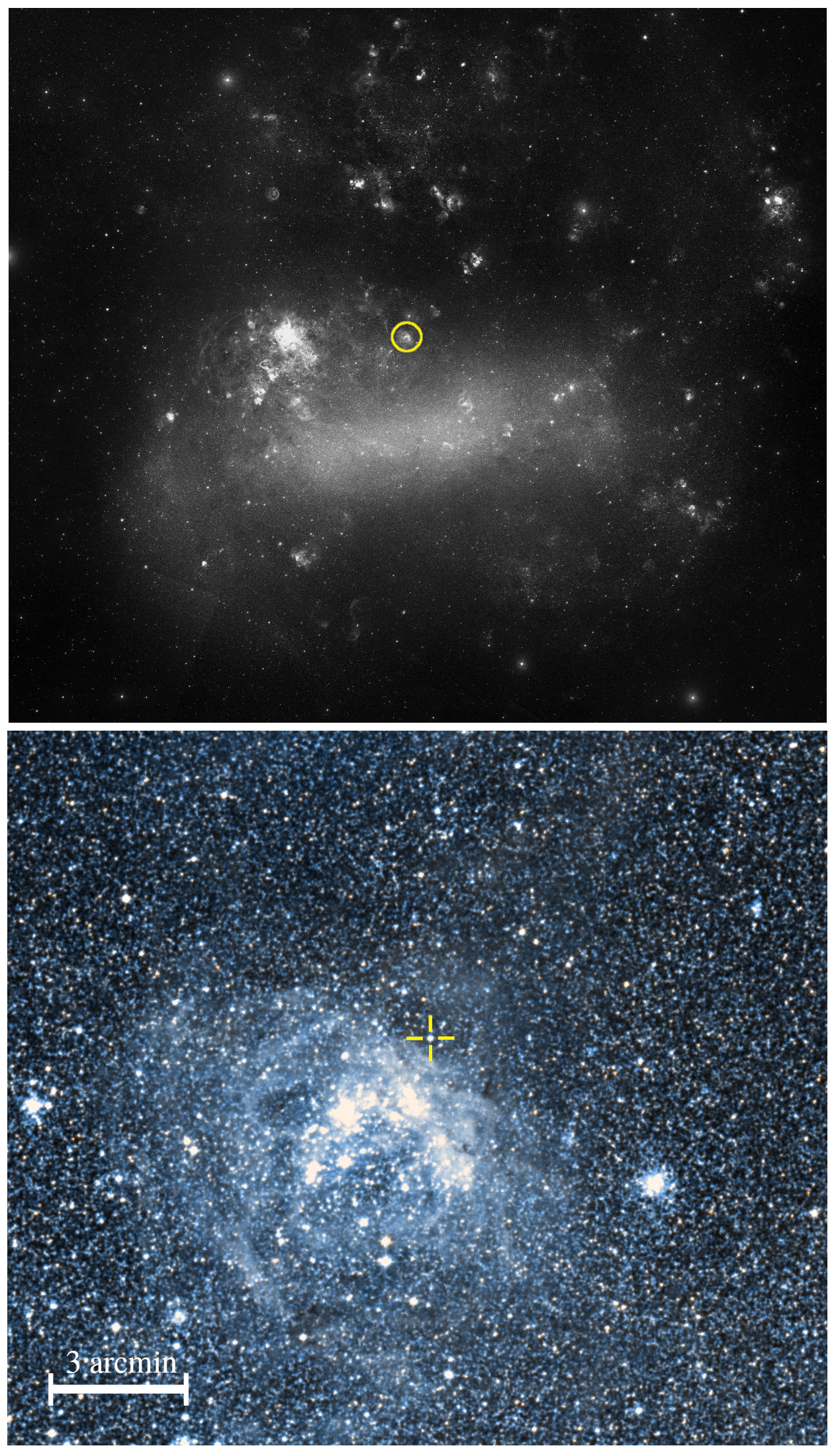}
   \caption{Top: Wide-field (8.0$\degr\,\times\,7.1\degr$) Digital Sky Survey (DSS) image showing the overall location of the ExtEEV in the LMC. The bright encircled object marks the position of the OB association with the ExtEEV located near it. Bottom: DSS coloured image showing the close-up view of the LMC around the ExtEEV (marked with a yellow cross). The bright nebula south-east of the ExtEEV is the OB association LH\,58. The images were generated using the Centre de Donn\'ees astronomiques de Strasbourg (CDS) \texttt{Aladin Lite} tool. In both images, north is up and east is to the left.}
   \label{fig:SDSS-view}
\end{figure}

The main obstacle in extracting the photometry of the ExtEEV from FFIs is the crowded field of the LMC (Fig.\,\ref{fig:SDSS-view}). Given the brightness of our target and the severe contamination caused by the low spatial resolution, it is difficult to perform reliable photometry of the ExtEEV using original FFIs. In order to overcome this problem, we decided to make photometry with difference images. They were obtained by means of the difference image analysis (DIA) on TESS FFIs using \texttt{pyDIA}\footnote{https://github.com/MichaelDAlbrow/pyDIA} open-source software \citep{2017zndo....268049M}. The mathematical part of \texttt{pyDIA} incorporates the formalism of DIA developed by \cite{2013MNRAS.428.2275B}. The main advantages of the \texttt{pyDIA} code are the use of the extended $\delta$-basis functions and taking the varying photometric scale across the image into account, which is especially important for ground-based observations with a large field of view.

\begin{figure*}
   \centering
   \includegraphics[width=0.8\hsize]{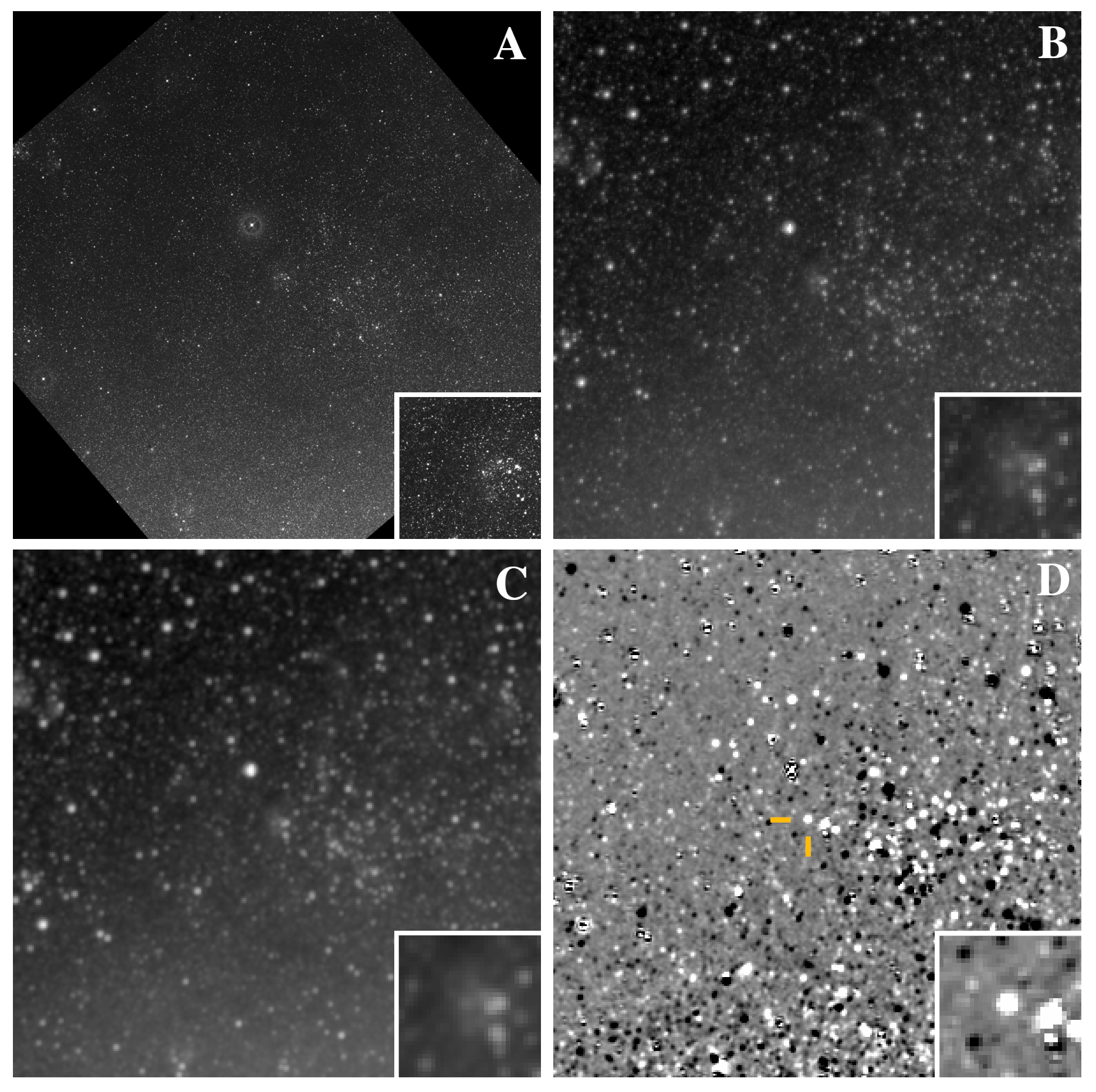}
   \caption{Summary of DIA analysis performed on TESS FFIs. Each panel is a $105\,\arcmin\,\times\,105\arcmin$ square centred on the position of the ExtEEV. The insets in the lower right corners show a zoomed view of the central areas, each $12\,\arcmin\,\times\,12\arcmin$. (A) DSS infrared image scaled and oriented as the TESS FFI cut-outs presented in panels B, C, and D. The image was generated using DSS. (B) A sample original 300\,$\times$\,300 pix cut-out of TESS FFI from sector 7. (C) The same as in panel B, but convolved with a 2D Gaussian kernel having constant standard deviation of $\sigma=$\,0.5 pix. (D) The difference image generated by \texttt{pyDIA} which corresponds to the image shown in panel C. The intersection of orange markers denotes the position of the ExtEEV. The images in panels B\,--\,D correspond to the epoch of maximum light of the ExtEEV.}
   \label{fig:DIAcomparison}
\end{figure*}

Using \texttt{pyDIA}, we extracted the light curve of the ExtEEV as follows. First, we cut out 300\,$\times$\,300 pix rasters from the original FFIs having a size of 2136\,$\times$\,2078 pix. Whenever possible, we placed the ExtEEV in the centre of a raster (Fig.\,\ref{fig:DIAcomparison}, panel B), but in several sectors, our object was too close to the edge of the image to meet this condition. \texttt{pyDIA} (and DIA in general) is not well suited to work with undersampled TESS FFIs. Hence, before we ran \texttt{pyDIA}, we convolved rasters with a spatially constant two-dimensional (2D) Gaussian with $\sigma=0.5$\,pix. The price was a slightly reduced resolution (Fig.\,\ref{fig:DIAcomparison}, panel C), but this allowed us to obtain difference images of a noticeably better quality. Finally, a series of difference images was generated using second-degree polynomials describing both the background changes and the spatial variation of the convolution kernel. The changes in the photometric scale were not allowed because TESS images were taken from space. In addition, in order to model noise in \texttt{pyDIA} fits, we used error images that were delivered by the TESS team. Before using these error images, we convolved them with the same 2D Gaussian that was used to smooth the science images. Panel D in Fig.\,\ref{fig:DIAcomparison} presents a representative difference image obtained in this procedure.

Having obtained the complete set of difference images, we performed non-circular aperture photometry of the ExtEEV, taking the subtraction of the residual background into account, which was estimated as the median of the signal in the surrounding pixels lying within the background mask. The resulting light curve was cleared of obvious outliers using the iterative $\sigma$-clipping method. In general, our method of extracting the TESS light curve of the ExtEEV differed from that of Jaya21. These authors also applied DIA, but they used the ISIS package \citep{1998ApJ...503..325A,2000A&AS..144..363A} and its adaptation to the TESS FFIs developed by \cite{2021MNRAS.500.5639V}. Our approach mainly differs in that we used smaller FFI cut-outs for the DIA and performed aperture photometry, while they applied profile photometry.

\begin{figure}
   \centering
   \includegraphics[width=\hsize]{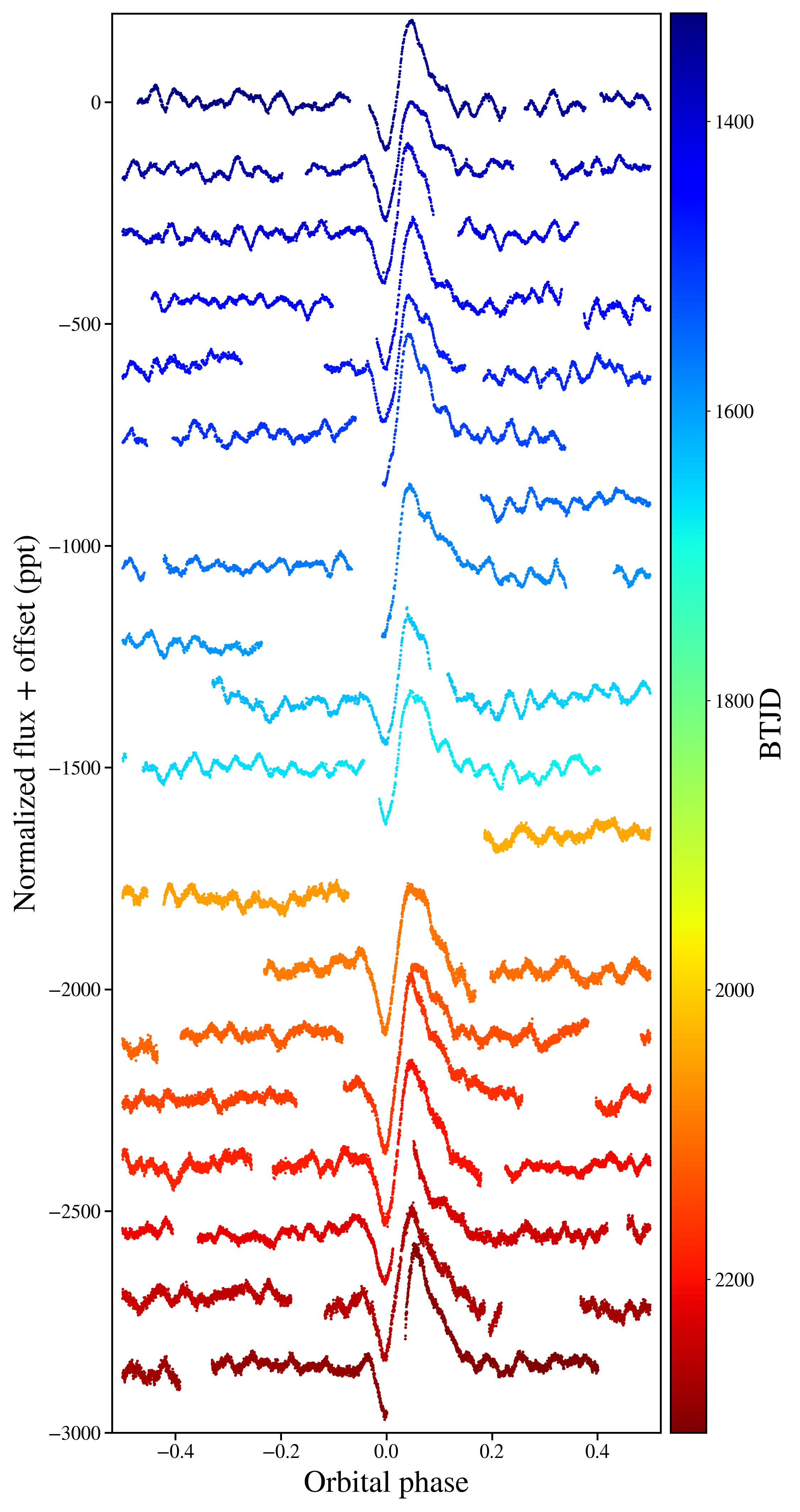}
   \caption{Phased light curve of the ExtEEV obtained by means of \texttt{pyDIA}. The consecutive orbital cycles are vertically shifted by 150\,ppt in order to present a non-repetitive character of the heartbeat and the other variability. Here and throughout our paper, BTJD\,$\equiv$\,BJD$-$2457000. The zero phase corresponds to the time of minimum brightness of the heartbeat, BJD\,2458373.61518. The light curve was phase-folded with the orbital period of 32.83016\,d.}
   \label{fig:tess-lc}
\end{figure}

The last step necessary to obtain the light curve expressed in relative flux units was to determine the reference flux of the ExtEEV in reference images returned by \texttt{pyDIA}. The aperture necessary to measure the total flux of the ExtEEV in the TESS reference image has a diameter of approximately 6 pixels, which corresponds to about 2$\arcmin$ in the sky. This means that the flux measured in TESS images suffers from severe contamination from nearby stars. We attempted to estimate this contamination in the following way. We downloaded the $I$-band Optical Gravitational Lensing Experiment (OGLE) (Sect.~\ref{sect:ogle}) reference image of the area containing the ExtEEV. The spatial resolution of OGLE-III images amounts to $\sim$0.26\,$\arcsec$/pix. Therefore, the ExtEEV is isolated well from other stars in the OGLE-III frames. By convolving the aforementioned OGLE-III image with a spatially constant 2D Gaussian function, we estimated the contamination of the ExtEEV in the TESS FFIs for approximately 90\%. In other words, only about 10\% of the flux we measured in the TESS difference images may actually come from the ExtEEV. Unfortunately, after applying the estimated correction, we were not able to reproduce the peak-to-peak amplitude of the heartbeat observed in the OGLE data. Eventually, we decided to convert the TESS light curve to relative flux units assuming that the average range of brightness changes due to the heartbeat in the OGLE-III/-IV and TESS data is the same. Figure \ref{fig:tess-lc} shows the whole TESS light curve of the ExtEEV which entered our subsequent analysis. The total time span of the light curve amounts to approximately 2.7 yr with a central $\sim$1-year long gap. The presence of this gap results in the occurrence of yearly aliases in the Fourier frequency spectra of TESS data, which can be seen, for instance, in Figs.\,\ref{fig:tess-peri-all}d and e. 

\subsection{The heartbeat}\label{sect:the-heartbeat}
It can be seen in Fig.\,\ref{fig:tess-lc} that the shape of the heartbeat is not exactly repeatable during the successive periastron passages. The reason for these changes is difficult to explain without time-resolved spectroscopy and modelling of the photometric yield of the system. It can only be speculated that a significant role in these processes must be played by the primary's stellar wind and dispersed matter, for instance in the form of a fast-evolving disk.

In addition to the well pronounced variability at the periastron, smaller variability with a total range of about 100\,ppt can be seen throughout the whole TESS light curve, especially outside the periastron. These changes originate from the superposition of several coherent TEOs (Sect.~\ref{sect:global-pre-whitening}) and additional stochastic variability which we discuss in Sect.~\ref{sect:stochastic}. It cannot be ruled out either that some part of the stochastic variability observed in the ExtEEV is of non-stellar origin and has its source in a disk, which may surround one or both components. Its size may vary as a function of the orbital phase, as suggested by Jaya21.

\subsection{Analysis of the TESS light curve}\label{sect:global-pre-whitening}
\begin{figure*}
   \centering
   \includegraphics[width=0.9\hsize]{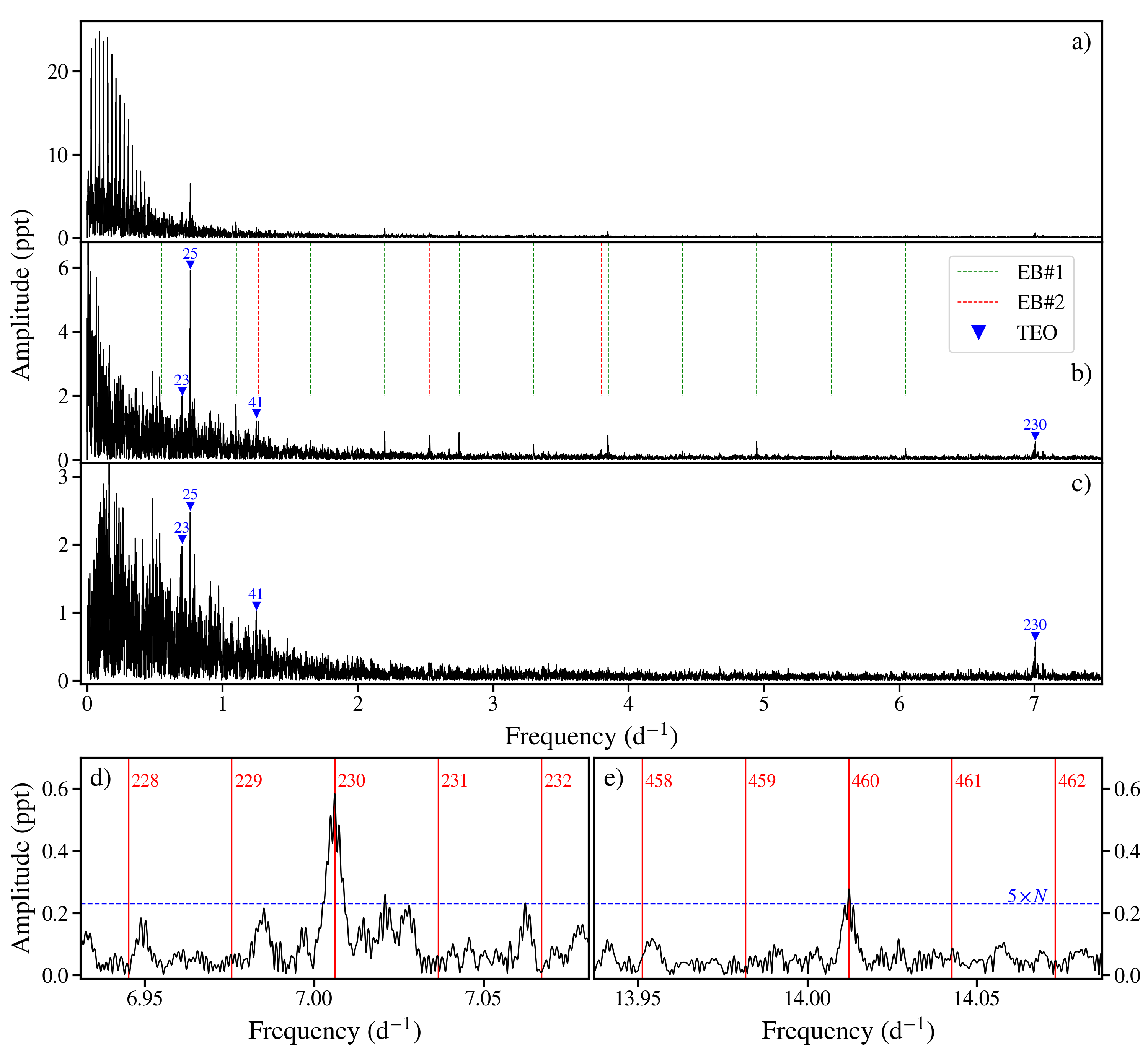}
   \caption{Fourier frequency spectra of the TESS light curve of the ExtEEV. (a) Frequency spectrum of the original TESS light curve. (b) The same as in (a), but after subtracting the heartbeat signal contributing to the lowest ($n=1$ to 17) orbital harmonics. Vertical dashed lines indicate frequencies present due to the contamination by the two nearby eclipsing binaries, EB\#1 (green) and EB\#2 (red, only odd harmonics). More details are provided in the main text. Blue triangles mark the position of the detected TEOs and are labelled with $n$. (c) The same as in (b), but after subtraction of the signal from EB\#1, EB\#2, and the dominant $n=25$ TEO. A mild detrending at low frequencies was also applied. (d) and (e) Zoomed parts of the frequency spectrum shown in panel (c), in the vicinity of $n=230$ TEO (d) and its lowest harmonic (e). Red vertical lines denote the position of the consecutive orbital frequency harmonics labelled with $n$. The dashed horizontal line shows the detection level defined as five times the noise in the spectrum calculated in the range 10\,--\,40\,{\cd}.}
   \label{fig:tess-peri-all}
\end{figure*}

We started our analysis by calculating the Fourier frequency spectrum of the whole TESS light curve (Fig.\,\ref{fig:tess-peri-all}a). As expected, the spectrum is dominated by the orbital frequency and its low ($n=2$ to 17) harmonics, which can be seen as the comb-like structure at low frequencies. Their occurrence is an obvious result of the domination of the variability due to the heartbeat in the light curve (Fig.\,\ref{fig:tess-lc}). The orbital period derived from all TESS data is equal to 32.83016\,$\pm$\,0.00011\,d. Having subtracted these harmonics, we get the frequency spectrum shown in  Fig.\,\ref{fig:tess-peri-all}b with an enhanced signal at low frequencies, resembling red noise. The signal originates from stochastic variability common among hot main-sequence stars and early-type supergiants. We discuss this component of variability in detail in Sect.~\ref{sect:stochastic}. In addition to the stochastic signal, some superimposed distinct peaks can be seen in the spectrum. We identified these peaks in the following way. For frequencies higher than 2.5\,{\cd}, we considered a peak to be  statistically significant if its height exceeded five times the noise level, which was calculated as an average signal in the range of 10\,--\,40\,{\cd}. This frequency range is free from the influence of the stochastic variability seen at low frequencies. Below 2.5\,{\cd}, we adopted only those frequencies that were close to the harmonics of the orbital frequency or those that were sub-harmonics of significant frequencies located above 2.5\,{\cd}. The analysis included a standard iterative pre-whitening procedure, using non-linear least-squares to fit a truncated Fourier series.

\begin{figure}
   \centering
   \includegraphics[width=\hsize]{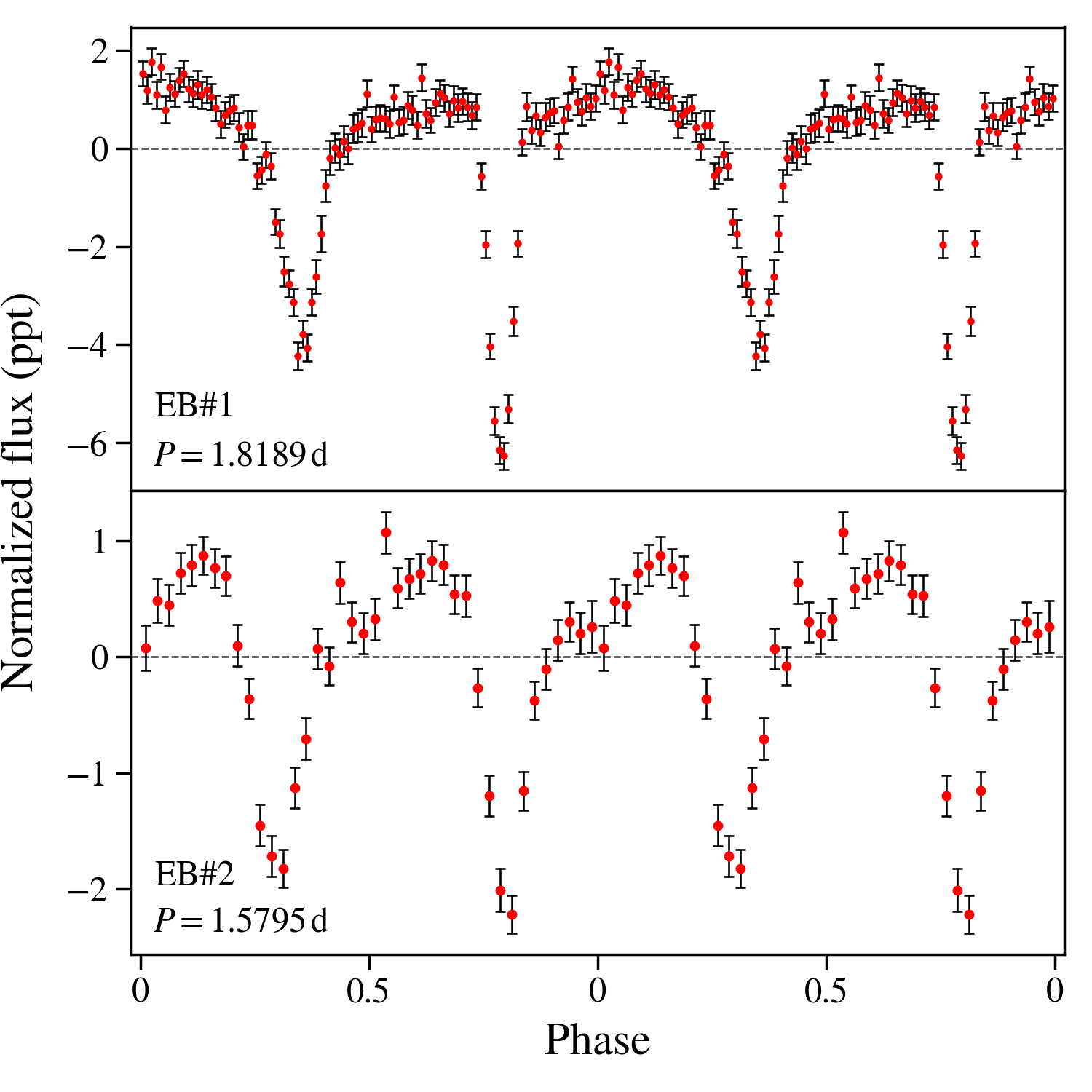}
   \caption{Phased light curves of two eclipsing binaries separated from our TESS light curve of the ExtEEV. Each point represents the median value calculated in the 0.01 (upper) and 0.025 (lower) phase bins. The upper panel shows the light curve of OGLE-LMC-ECL-15664, and the lower shows that of OGLE-LMC-ECL-15785 (HD\,269548). Phase 0.0 corresponds to BJD\,2458325.6366 (for EB\#1) and BJD\,2458325.7898 (for EB\#2).}
   \label{fig:EBs}
\end{figure}

Figure \ref{fig:tess-peri-all}b shows that there are some peaks at high frequencies that are harmonics of two frequencies, $\sim$0.550\,{\cd} and $\sim$1.266\,{\cd}. Therefore, we checked a possible contamination by nearby variable stars. Using the OGLE catalogue of eclipsing binaries \citep{2016AcA....66..421P}, we found that these signals occur as a result of contamination by two nearby eclipsing binaries, OGLE-LMC-ECL-15664 = MACHO 80.7443.1746 (EB\#1, $P_{\rm orb}=1.8189$\,d), separated by 27$\arcsec$ from the ExtEEV, and OGLE-LMC-ECL-15785 = MACHO 80.7564.37 (HD\,269548, EB\#2, $P_{\rm orb}=1.5796$\,d), separated by 88$\arcsec$ from the ExtEEV. Phased light curves of these two eclipsing binaries extracted from our TESS light curve (Fig.\,\ref{fig:EBs}) match the morphology of their light curves obtained by the OGLE team. Some differences for EB\#1 can be explained by the apsidal motion in this system \citep{2020A&A...640A..33Z}. Both systems are relatively bright ($V =$\,15.8 and 15.6\,mag, respectively), which explains why they contaminate the light curve of the ExtEEV.

\begin{figure}
   \centering
   \includegraphics[width=\hsize]{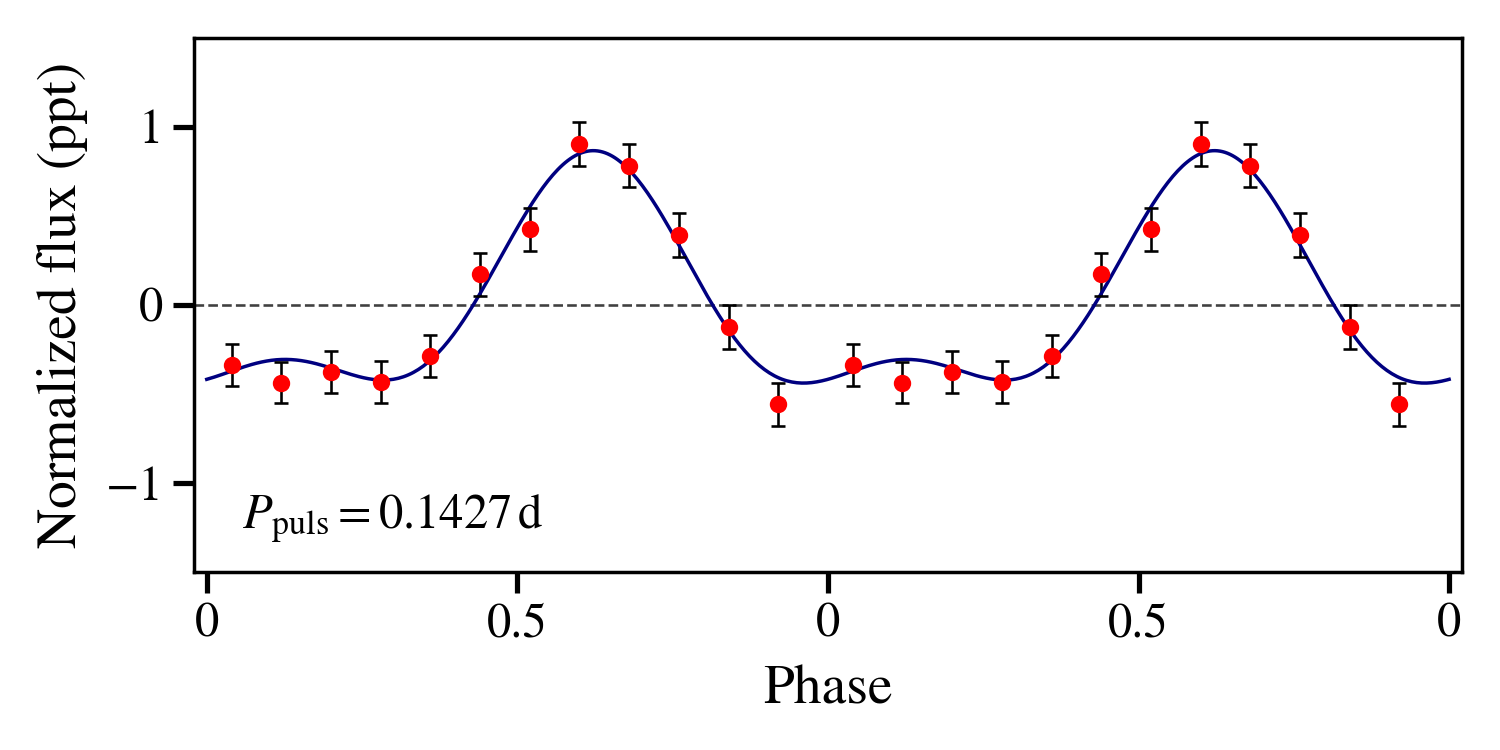}
   \caption{TESS light curve of $n=230$ TEO phased with its pulsation period. Red circles denote the median flux calculated in 0.08 phase bins, and the dark blue line corresponds to the truncated Fourier series consisting of the pulsation frequency and its lowest harmonic.}
   \label{fig:n230-phased-lc}
\end{figure}

It can already be seen in Fig.\,\ref{fig:tess-peri-all}a, but is evident from Fig.\,\ref{fig:tess-peri-all}b, that a large-amplitude $n=25$ TEO is present in the light curve of the ExtEEV. It was detected by Jaya19 and recently confirmed by Jaya21. After subtracting this TEO and applying some mild detrending which removed the signal at the lowest frequencies, we obtained the frequency spectrum which is shown in Fig.\,\ref{fig:tess-peri-all}c. The spectrum shows three additional TEOs at $n=23, 41$, and 230. The $n=41$ TEO was already found by Jaya21. These authors also indicated the possibility of the presence of the $n=23$, but it appeared in the frequency spectrum of the light curve confined to out-of-heartbeat phases, so it could have been an orbital alias. The $n=230$ TEO at the frequency 7.006\,{\cd} is a new finding. This high-$n$ TEO has its own harmonic at the frequency 14.012\,{\cd} (Fig.\,\ref{fig:tess-peri-all}e), which results in an unusual light curve (Fig.\,\ref{fig:n230-phased-lc}). We estimated the probability that this TEO is a self-excited mode in some contaminating star or the secondary, occurring by chance near the harmonic of the orbital frequency. Taking the error of the orbital frequency into account, $\sigma_{f_{\rm orb}} = 1.02\,\times\,10^{-7}$\,{\cd}, the probability of detecting a self-excited mode with random frequency within $\pm 3\sigma_{f_{\rm orb}}$ around $n={\rm 230}$ can be estimated as equal to $6\,\times\,230\,\times\,\sigma_{f_{\rm orb}}/f_{\rm orb}\approx4.6\,\times\,10^{-3}$. Hence, the probability is very low, which can also be judged by comparing the separation of the harmonics with the width of peaks in Figs.\,\ref{fig:tess-peri-all}d and e. The frequency of 7.006\,{\cd} is neither a harmonic of EB\#1 or EB\#2 nor do the OGLE $I$-band light curves show coherent variability with this frequency. We therefore conclude that it is a bona fide TEO. The $n=230$ TEO is discussed in detail in Sect.~\ref{sect:why-teo-230}.

Another surprising feature in Fig.\,\ref{fig:tess-peri-all}c is the residual signal in the vicinity of the $n=25$ TEO. This may indicate the amplitude or phase change in this TEO and prompted us to carry out a detailed analysis of the changes in the amplitudes and periods of TEOs seen in the ExtEEV, which is presented in Sect.\,\ref{sect:changing-a-and-f}. Table \ref{table:global-pre-whitening} summarises the results of the analysis carried out in the present section. The presented frequencies and amplitudes were obtained by means of the least-squares fitting to the whole TESS data set and, therefore, they represent some average values.
\begin{table*}
\caption{Periodic signals present in the TESS light curve of the ExtEEV.}
\label{table:global-pre-whitening}      
\centering                         
\begin{tabular}{r@{.}l c r r@{.}l l}        
\hline\hline                 
\noalign{\smallskip}
\multicolumn{2}{c}{Frequency\,(\cd)} & Amplitude\,(ppt) &$n^a$ & \multicolumn{2}{c}{${\Delta n}^{a,b}$} & Comment \\
\noalign{\smallskip}
\hline  
\noalign{\smallskip}
0&03045980(10) & --  & -- & \multicolumn{2}{c}{--} & $f_{\rm orb}$, harmonics up to $n = 17$ detected\\
\noalign{\medskip}
0&70058(3) & 1.00(13) & 23 & $+$0&0002(10) & new TEO \\
0&762049(11) & 5.96(13)  & 25 & $+$0&0181(4) & known TEO\\
1&24940(6) & 1.12(13) & 41 & $+$0&0177(19) & known TEO \\
7&00612(8) & 0.59(13) & 230 & $+$0&0111(27) & new TEO \\
14&01223(16) & 0.28(13) & 460 & $+$0&024(5) & harmonic of $n=$\,230 TEO  \\
\noalign{\medskip}
0&549783(5) & -- & -- & \multicolumn{2}{c}{--} & $f_{\rm orb, EB\#1}$, 9 consecutive harmonics detected \\
0&63311(3) & -- & -- & \multicolumn{2}{c}{--} & $f_{\rm orb, EB\#2}$, only three lowest odd harmonics detected\\
\noalign{\smallskip}
\hline                                   
\end{tabular}
\tablefoot{$^a$ Specified only for TEOs. $^b$ $\Delta n=f/f_{\rm orb}-n$, where $f$ is the detected frequency (first column).}
\end{table*}

\subsection{Low-$n$ TEOs and rotation period}
Jaya19 claim the detection of two additional TEOs at $n=8$ in the ASAS-SN data and $n=7$ TEO in the TESS data. Because the ASAS-SN data are ground-based data, the former is clearly a daily alias of the $n=25$ TEO, because the sum of its frequency, 0.241\,$\pm$\,0.012\,{\cd}, and the frequency of the $n=25$ TEO, 0.761\,$\pm$\,0.007\,{\cd}, is equal to 1.003\,$\pm$\,0.014\,d$^{-1}$, that is, within the errors, 1 (sidereal day)$^{-1}$. An attempt to fit both frequencies resulted in a rather uncertain result (their table~3): amplitudes much lower than in the frequency spectrum and errors of all parameters much higher than expected. Concluding, the $n=8$ TEO is spurious. 

The other frequency, 0.2164\,$\pm$\,0.0004\,d$^{-1}$, detected with an amplitude of about 11\,ppt by Jaya19 in the TESS data, and tentatively assigned to $n=7$ TEO, was not confirmed by Jaya21. The latter authors claim, however, that another term with a similar frequency of 0.2253\,$\pm$\,0.0012\,d$^{-1}$ and an amplitude of 4.4\,$\pm$\,0.8\,ppt (their table~5) is significant and may represent the rotational period of the primary. Judging from the lower panel of their fig.\,11, it seems, however, that there are many other peaks with a similar signal-to-noise ratio and this one is by no means the highest. Our own analysis did not reveal any isolated and distinctive maximum close to this frequency, where the signal does not exceed 2.5\,ppt. It seems, therefore, that the two frequencies close to 0.22\,d$^{-1}$ found by Jaya19 and Jaya21 represent two of many peaks characteristic of the stochastic variability discussed in the next subsection and seen in Fig.\,\ref{fig:tess-peri-all}b. There is no good reason to assume that any of them represent the rotation frequency of the primary.

\subsection{Stochastic variability}\label{sect:stochastic}
The frequency spectrum calculated for the TESS residual light curve of the ExtEEV (Fig.\,\ref{fig:tess-peri-all}b) reveals a significant signal in the low-frequency range. It is a signature of stellar stochastic variability, which is typical for many massive early-type stars (see \cite{2019A&A...621A.135B,2019NatAs...3..760B,2020A&A...640A..36B} for a compilation of numerous examples). There are three main possible explanations of this variability. The first includes internal gravity waves as the likely source. Stars that are born with masses $\gtrsim$\,1.5\,M$_\odot$ have a convective core. At its boundary, the turbulent convection may excite both the coherent and damped stochastic oscillations that propagate towards the stellar surface \citep[e.g.][]{2013MNRAS.430.1736S,2021MNRAS.508..132L}. The second possible explanation is a thin subsurface convection layer which should manifest itself as a `flickering' granulation pattern on a stellar surface. Recently, \cite{2021ApJ...915..112C} have shown that properties of the subsurface convection zone driven by the iron opacity peak correlate with observational characteristics of low-frequency stochastic variability. Finally, non-uniform and clumpy stellar winds for which signatures are observed in the spectra of stars with masses higher than $\sim$15\,M$_\odot$ \citep[e.g.][]{2018A&A...617A.121K,2018MNRAS.473.5532R} may also contribute to this type of variability.

As was shown by \cite{2020A&A...640A..36B}, a low-frequency signal due to the stochastic variability can be described with the Lorentzian profile,
\begin{equation}
\label{eq:igw}
    \alpha(f)=\frac{\alpha_0}{1+(f/f_{\rm char})^\gamma}+C_{\rm w},
\end{equation}
where $f$ denotes frequency, $\alpha_0$ is the amplitude at $f=0$, $f_{\rm char}$ is the characteristic frequency which reflects the characteristic timescale of stochastic variability, $\gamma$ is the logarithmic amplitude gradient, and $C_{\rm w}$ accounts for the presence of the `white noise' in the frequency spectrum parallel to the `red noise' component. Using the Markov chain Monte Carlo (MCMC) sampler implemented in the Python \texttt{emcee} package \citep{2013PASP..125..306F}, we fitted the function given by Eq.\,(\ref{eq:igw}) to the frequency spectrum of the ExtEEV calculated at the residuals with heartbeat phases (between $-$0.05 and 0.2 in Fig.~\ref{fig:tess-lc}) being cut out  in order to exclude these parts from the analysis. We initialised 100 walkers each 100\,000 steps long, preceded by a burn-in phase of 500 steps. We used flat priors and `thinned' the final chains by half of the estimated correlation time, which was about 40 for all parameters.
\begin{figure}
   \centering
   \includegraphics[width=\hsize]{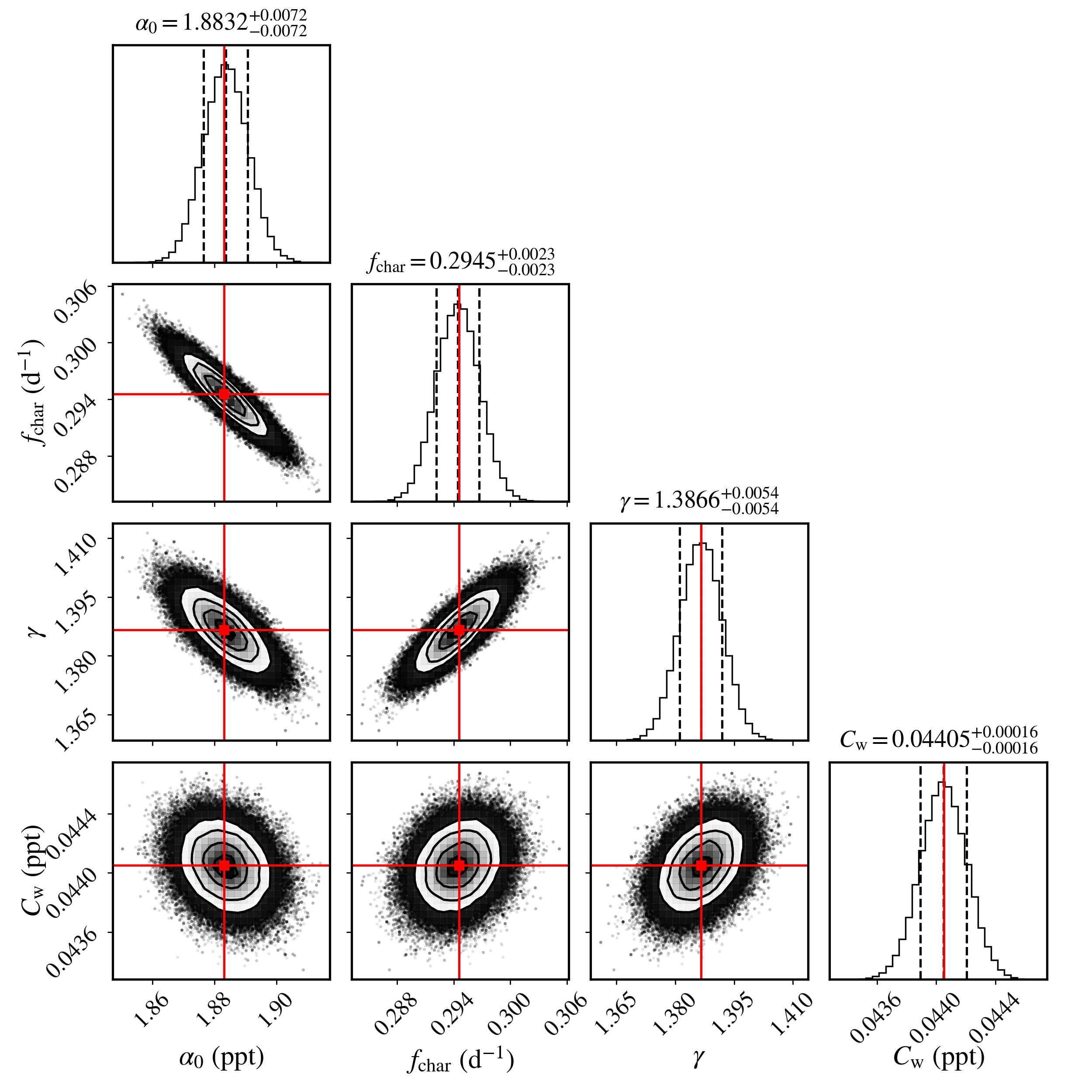}
   \caption{Corner plot resulting from our MCMC analysis of the TESS residual light curve of the ExtEEV. Red markers indicate the best-fit solution. Three vertical dashed lines superimposed on each marginalised one-dimensional posterior distribution denote 16\%, 50\%, and 84\% quantiles.}
   \label{fig:igw-cornerplot}
\end{figure}
\begin{table}
\caption{Best-fit parameters from  Eq.~(\ref{eq:igw}) resulting from the MCMC analysis of the residual frequency spectrum of the ExtEEV. See the main text for an explanation of the parameters.}
\label{table:igw-params}      
\centering                         
\begin{tabular}{c l}        
\hline\hline                 
\noalign{\smallskip}
Parameter & Value\\    
\noalign{\smallskip}
\hline  
\noalign{\smallskip}
   $\alpha_0$ (ppt) & 1.883(7)\\
   $f_{\rm char}$ (\cd) & 0.2945(23)\\
   $\gamma$ & 1.387(5)\\
   $C_{\rm w}$ (ppt) & 0.04405(16)\\
\noalign{\smallskip}
\hline                                   
\end{tabular}
\end{table}
\begin{figure}
   \centering
   \includegraphics[width=\hsize]{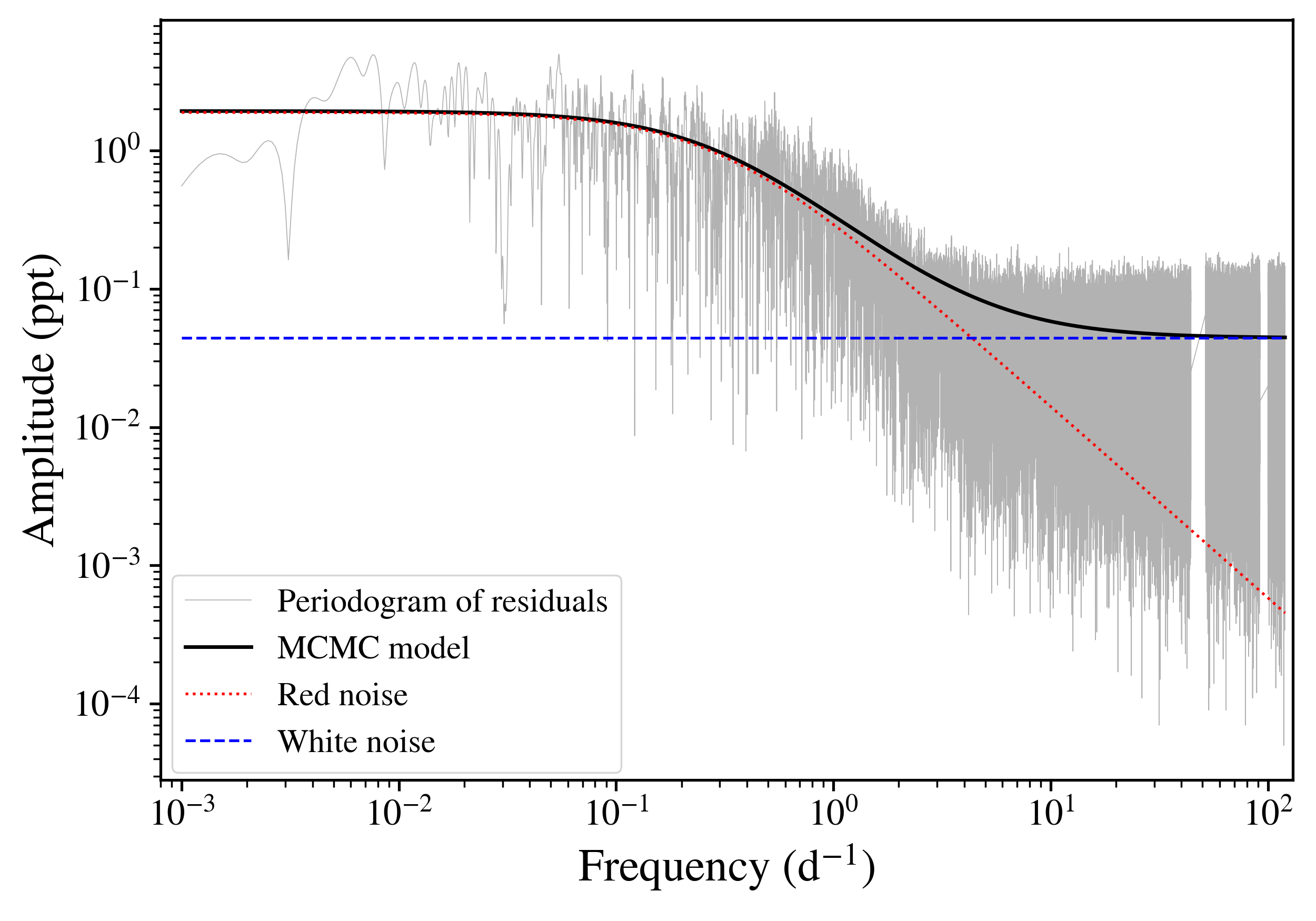}
   \caption{Frequency spectrum of the residual TESS light curve after subtracting all significant coherent signals and rejecting data close to the periastron passages. The black solid line represents the best-fit model given by Eq.~(\ref{eq:igw}). The stochastic component of the model is denoted with a red dotted line while the horizontal blue dashed  line corresponds to the white noise level. The two gaps in the data above 10\,d$^{-1}$ correspond to the location of aliases around the even multiples of the Nyquist frequency that we removed before the fit.}
   \label{fig:igw-fit}
\end{figure}

Figure \ref{fig:igw-cornerplot} shows the corner plot resulting from the MCMC simulation we performed while Table \ref{table:igw-params} presents the best-fit parameters. The residual frequency spectrum with the superimposed best-fit of Eq.\,(\ref{eq:igw}) can be seen in Fig.\,\ref{fig:igw-fit}. The enhanced signal at low frequencies stabilises at the level of white noise only at a frequency of about 10\,d$^{-1}$. In order to compare the behaviour of stochastic variability in the ExtEEV with those observed in other massive stars, we placed them in the analogue of the so-called spectroscopic Hertzsprung-Russell diagram (sHRD) presented by \cite{2020A&A...640A..36B} (their fig.\,2). 
\begin{figure*}
   \centering
   \includegraphics[width=0.9\hsize]{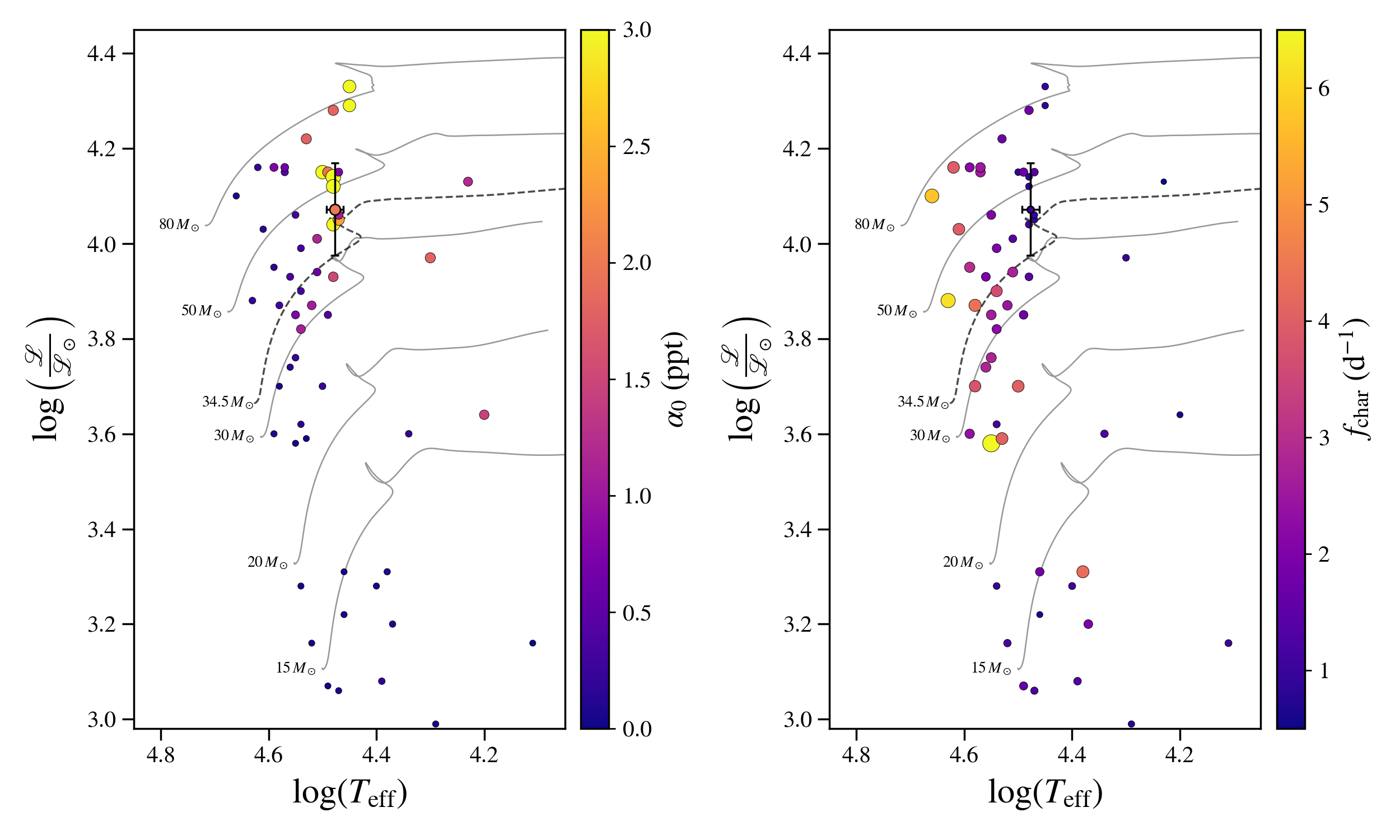}
   \caption{Spectroscopic HR diagrams showing the location of massive stars exhibiting low-frequency stochastic variability and analysed by \cite{2020A&A...640A..36B}. The spectroscopic luminosity is defined as $\mathscr{L}\equiv T_{\rm eff}^4/g$, where $T_{\rm eff}$ and $g$ denote the effective temperature and surface gravity of a star, respectively. The circles with error bars mark the position of the ExtEEV using its physical properties estimated by Jaya21. The colour and size of a symbol reflects the value of $\alpha_0$ (left) and $f_{\rm char}$ (right). Solid lines show MIST [Fe/H]$\,=\,$0, $v/v_{\rm crit}=0$ evolutionary tracks \citep{2016ApJ...823..102C,2016ApJS..222....8D} starting at the zero-age main sequence (ZAMS). The dashed-line evolutionary track corresponds to the suspected mass of the ExtEEV, 34.5\,M$_\odot$, assuming a mean metallicity of stars in the LMC, [Fe/H]$\,=-\,$0.4 and $v/v_{\rm crit}=0.4$.}
   \label{fig:igw-hr}
\end{figure*}
It can be clearly seen from Fig.\,\ref{fig:igw-hr} that both the amplitude $\alpha_0$ and characteristic frequency of stochastic changes $f_{\rm char}$ in the ExtEEV fit the overall trend visible in the sample analysed by \cite{2020A&A...640A..36B} very well. The more luminous and evolved a star is, the higher $\alpha_0$ is and the lower the value for $f_{\rm char}$ is. Therefore, the derived characteristics of the stochastic variability exhibited by the ExtEEV independently confirm a primary's evolutionary status: the star has already depleted hydrogen in its core or is very close to this phase. We found no evidence for a changing character of the stochastic variability before and after the periastron passage which means that whatever mechanism is responsible for this variability, it is not vulnerable to strong tidal deformation or it rebuilds quickly after the periastron passage.


\section{Changing amplitudes and frequencies of TEOs in the TESS data}\label{sect:changing-a-and-f}

\subsection{Separate analysis of two years of TESS data}\label{sect:TESS-separate-prewhitening}
\begin{figure*}
   \centering
   \includegraphics[width=0.9\hsize]{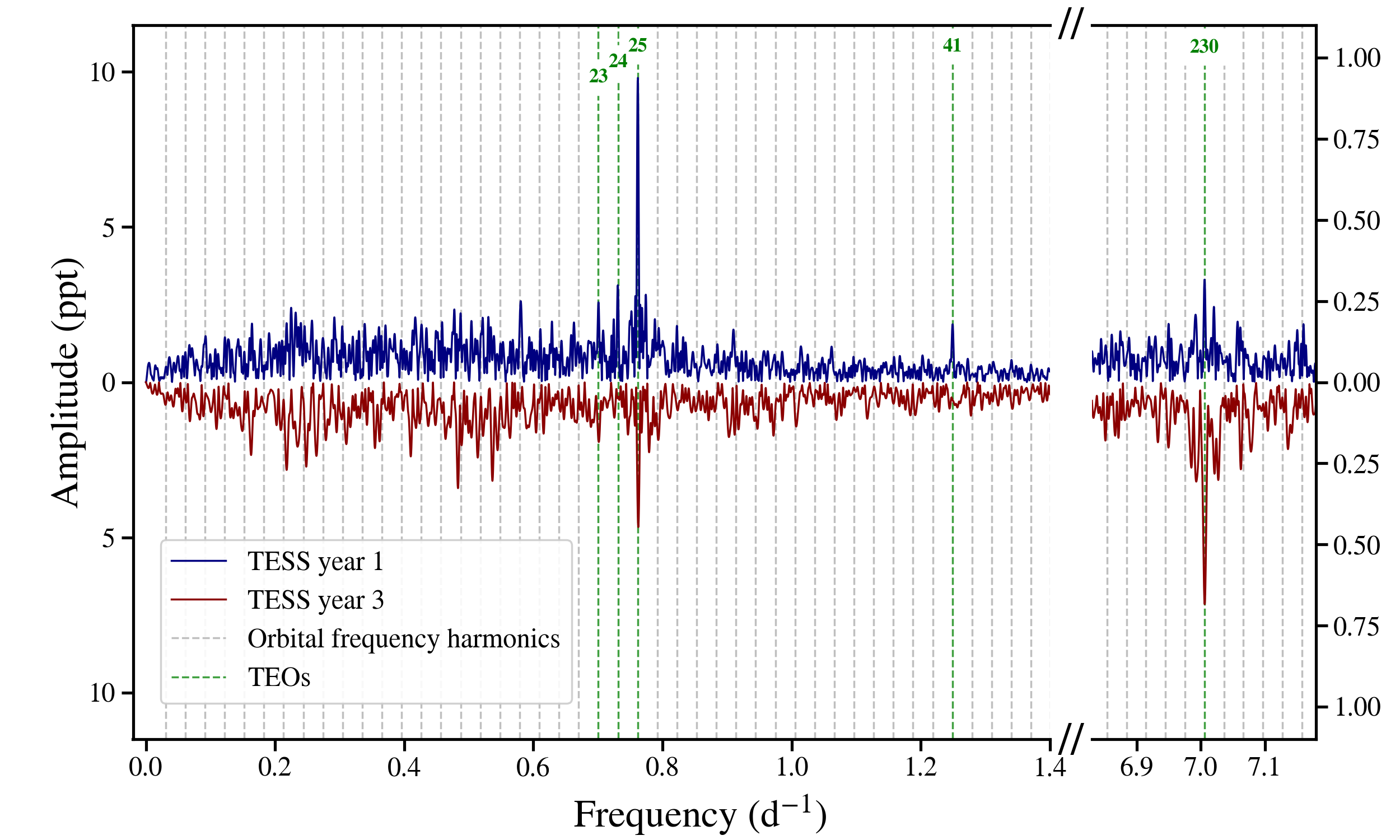}
   \caption{Comparison between Fourier frequency spectra calculated separately for TESS year 1 (upper blue curve) and TESS year 3 (lower red curve) data. For the sake of clarity, all variability, except TEOs, has been subtracted from the light curve. Detrending at the lowest frequencies was also applied. The vertical grey dashed lines mark the position of the consecutive harmonics of the orbital frequency. The location of detected TEOs is denoted with vertical green dashed lines and is labelled with $n$. Both parts of the frequency spectra have a different ordinate scale.}
   \label{fig:tess-2peri-comparison}
\end{figure*}

As we already noted in Sect.\,\ref{sect:global-pre-whitening}, there is an indication of changing amplitudes and/or periods of the TEOs in the target star. In order to verify this, we decided to apply the pre-whitening procedure separately for TESS year 1 and year 3 observations of the ExtEEV. Figure \ref{fig:tess-2peri-comparison} presents the comparison between Fourier frequency spectra calculated for these two groups of data. It can be clearly seen that all detected TEOs have different amplitudes in the two data sets. Without a doubt, the dominant $n=$\,25 TEO reduced its amplitude more than twice. On the contrary, the $n=230$ TEO increased its amplitude more than twofold. Some smaller changes in the amplitude of the $n=$\,41 TEO can also be seen. Surprisingly, the frequency spectrum calculated for TESS year 1 data reveals the presence of $n=24$ TEO, which is not detected in the TESS year 3 data. Furthermore, $n=23$ TEO seems to vanish in the TESS year 3 data. Table \ref{table:separate-pre-whitening} provides a quantitative description of these changes.

\begin{table}
\caption{Results of separate fits of TEOs for TESS year 1 and year 3 data. For each TEO, the first line corresponds to TESS year 1 data, and the second corresponds to TESS year 3 data.}
\label{table:separate-pre-whitening}    
\centering                         
\begin{tabular}{l r@{.}l c r@{.}l}
\hline\hline                 
\noalign{\smallskip}
Frequency\,(\cd) & \multicolumn{2}{c}{Amplitude\,(ppt)} & $n$ & \multicolumn{2}{c}{${\Delta n}^a$} \\
\noalign{\smallskip}
\hline  
\noalign{\smallskip}
0.70028(12) & 2&33(20) & 23 & $-$0&009(4) \\
not detected & $<$ 1&59 & 23& \multicolumn{2}{c}{---}\\
\noalign{\medskip}
0.73104(9) & 3&17(20) & 24 & $+$0&0005(30) \\
not detected & $<$ 0&40 &24& \multicolumn{2}{c}{---}\\
\noalign{\medskip}
0.76174(3) & 9&55(20) & 25 & $+$0&0081(11) \\
0.76246(7) & 4&52(16) & 25 & $+$0&0316(24) \\
\noalign{\medskip}
1.24893(15) & 1&95(20) & 41 & $+$0&003(5) \\
1.2491(4) & 0&80(16) & 41 & $+$0&009(14) \\
\noalign{\medskip}
7.0062(5) & 0&33(20) & 230 & $+$0&017(18) \\
7.0060(4) & 0&74(16) & 230 & $+$0&009(12) \\
\noalign{\smallskip}
\hline                                   
\end{tabular}
\tablefoot{$^a$ $\Delta n=f/f_{\rm orb}-n$, where $f$ is the detected frequency (first column).}
\end{table}

\begin{figure}
   \centering
   \includegraphics[width=\hsize]{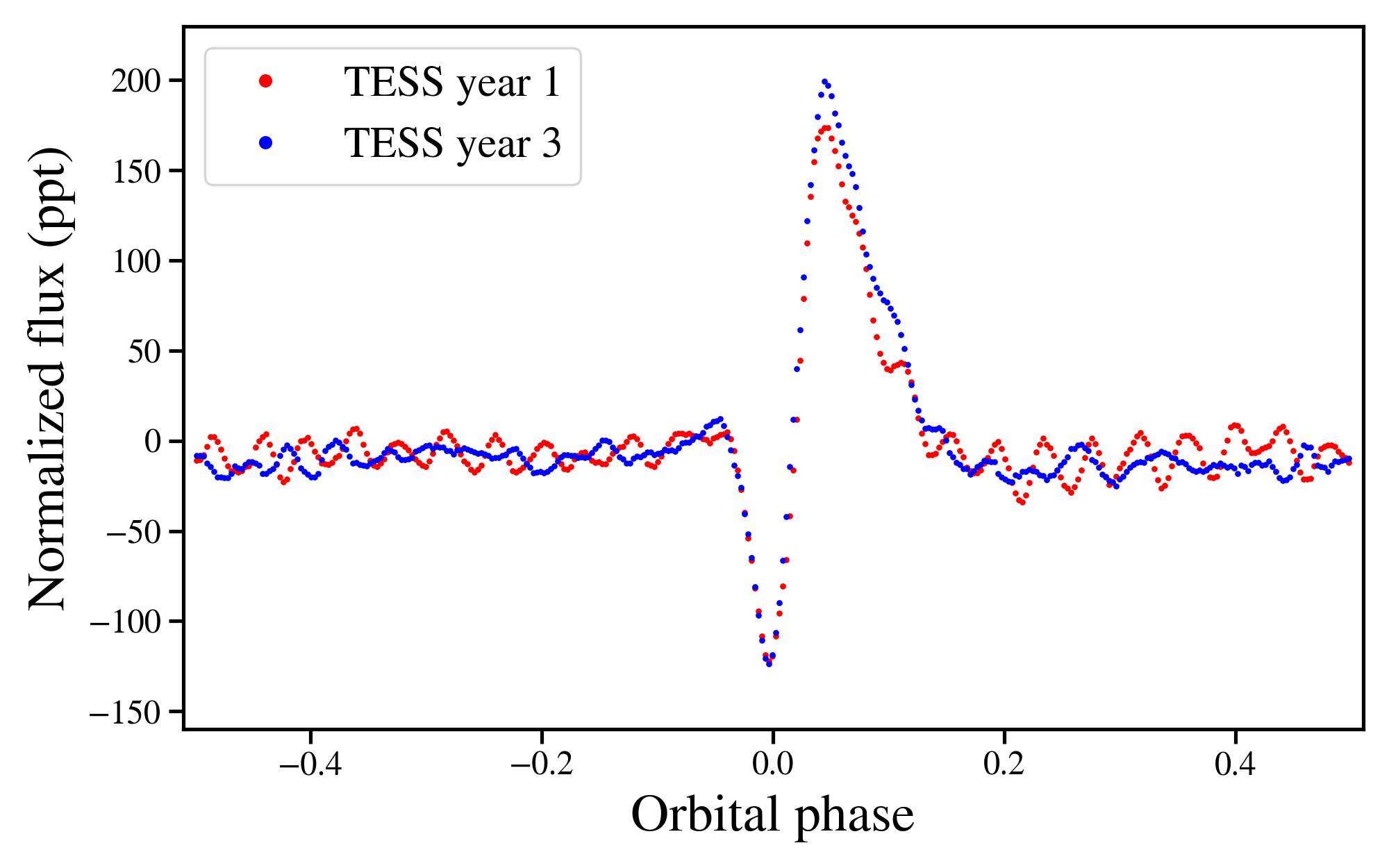}
   \caption{Comparison between phased TESS year 1 (red dots) and year 3 (blue dots) light curves of the ExtEEV, averaged in 0.003 phase bins. The variability, other than the heartbeat and TEOs, was subtracted.}
   \label{fig:tess-lc2}
\end{figure}

The change in amplitude of the strongest TEO can be clearly seen even in the phased light curve. Figure \ref{fig:tess-lc2} shows phased TESS year 1 and 3 light curves, freed from the contribution of variability other than the heartbeat and TEOs. It can easily be seen that the out-of-periastron fluctuations caused by TEOs are smaller in TESS year 3 data than in TESS year 1 data.

While a change in the amplitudes of TEOs between TESS year 1 and year 3 data is obvious from Fig.\,\ref{fig:tess-2peri-comparison} and Table \ref{table:separate-pre-whitening}, the centres of these two data sets are separated by about two years. We checked if the changes can be traced on a shorter timescale by analysing TESS data in shorter time intervals. The results of this analysis are given in the next subsection. 

\subsection{Changes in periods and amplitudes on a timescale shorter than one year}\label{sect:PW-TESS-sliding-window}
The analysis of amplitudes (and periods) of TEOs in time intervals shorter than one year has a clear limitation: the shorter the time interval, the higher the detection threshold. As a compromise between detectability and the possibility of tracing changes on as short a timescale as possible, we chose time intervals of the order of 100 d. In addition, the analysis was carried out only for the three strongest TEOs, $n=25$, 41, and 230.

The procedure was the following. In order to estimate the instantaneous amplitude of a TEO, we performed pre-whitening within the time `window' sliding along the TESS light curve. For the $n=25$ TEO, the window was 80\,d long, while the sliding step amounted to 27\,d. For $n = 41$ and 230 TEOs, we used 110\,d-long windows and a sliding step of 54\,d. We also checked the stability of the periods of the TEOs. For this purpose, we constructed ${O-C}$ diagrams. They were calculated from the times of maximum light derived by least-squares fitting of truncated Fourier series to the data in a window. The fits provided both the amplitudes and times of the maximum. The values of $O-C$ were calculated with respect to the ephemeris of the form $C = T_{\rm max} = T_0 + P_0\times E$, where $E$ is the number of cycles that elapsed from the initial epoch, $T_0$. The initial epochs and reference periods, $P_0$, are given in Table \ref{table:ephem}.
\begin{table}
\caption{Reference periods, $P_0$, and initial epochs, $T_0$ used to calculate $O-C$ diagrams shown in Fig.\ref{fig:teo-ampl-oc}.}
\label{table:ephem}      
\centering                         
\begin{tabular}{rcc}        
\hline\hline                 
\noalign{\smallskip}
$n$ & $T_0$ (BJD) & $P_0$ (d) \\
\noalign{\smallskip}
\hline  
\noalign{\smallskip}
25 & 2458325.633594 & 1.312775 \\
41 & 2458325.777057 & 0.800749 \\
230 & 2458325.457594 & 0.142737 \\
\noalign{\smallskip}
\hline                                   
\end{tabular}
\end{table}

\begin{figure}
   \centering
   \includegraphics[width=\hsize]{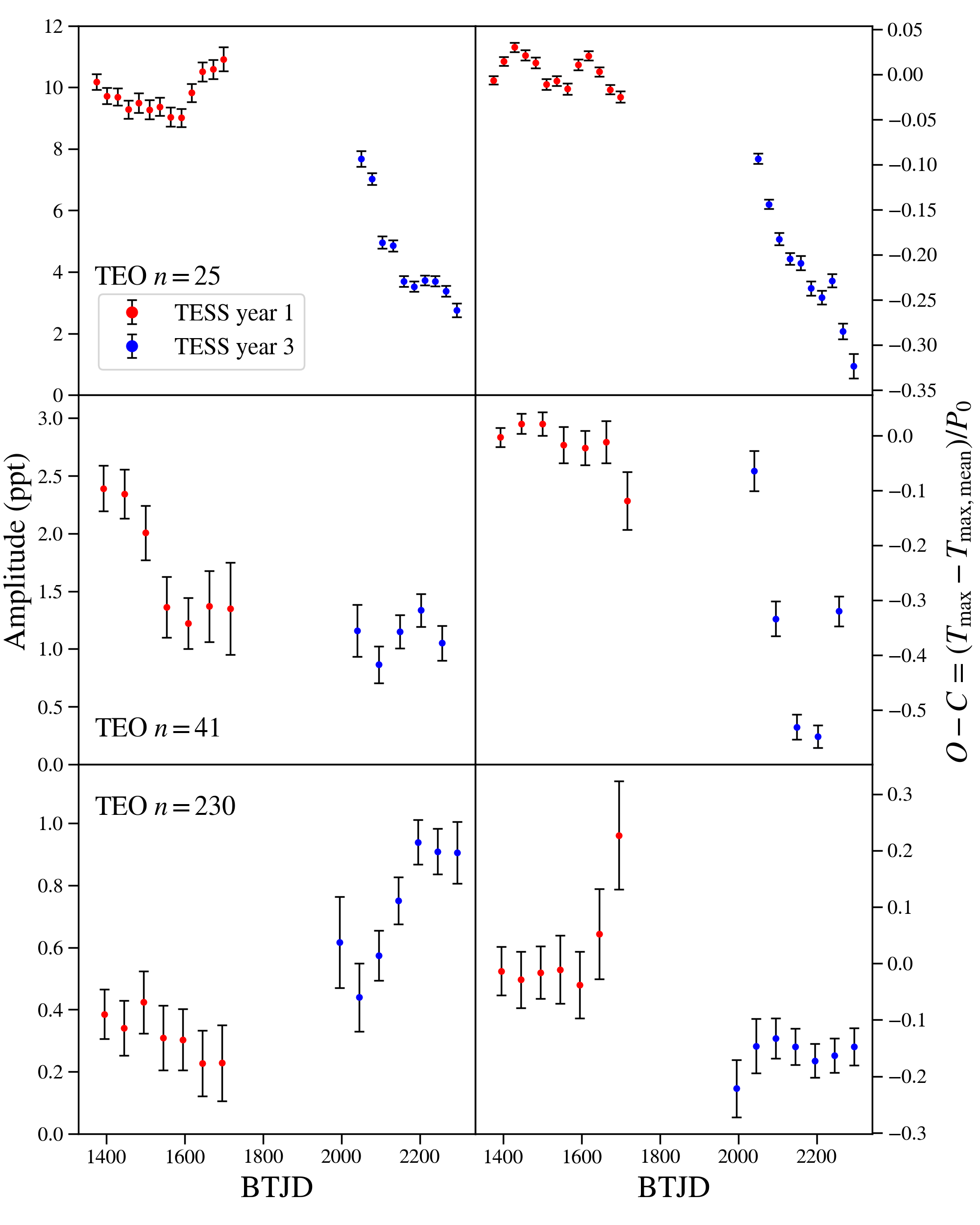}
   \caption{Amplitude changes (left) and $O-C$ diagrams (right) for TEOs with $n=$\,25, 41, and 230 detected in the TESS light curve. We note that $O-C$ values are expressed in the units of the period of the indicated TEO. See the main text for details.}
   \label{fig:teo-ampl-oc}
\end{figure}

The results of this analysis are summarised in Fig.\,\ref{fig:teo-ampl-oc}. The $n=25$ TEO exhibits by far the strongest relative changes in amplitude, reducing it from a maximum value of $\sim$11\,ppt to about 3\,ppt. The beginning of TESS year 3 observations indicates that it took about two months, that is, about two orbital cycles, for the amplitude to drop twice. Figure \ref{fig:teo-ampl-oc} also provides convincing evidence of the relatively fast and strong fluctuations of the amplitudes of $n=41$ and 230 TEOs. They confirm the picture of changes already seen in Fig.\,\ref{fig:tess-2peri-comparison}: a drop in amplitude of the $n=41$ TEO and an increase in amplitude of the $n=230$ TEO.

Alongside changing amplitudes, the $O-C$ diagrams for $n=25$ and 41 TEOs show statistically significant changes in the periods of these forced oscillations. According to Table \ref{table:separate-pre-whitening}, $n=25$ TEO increased its frequency between year 1 and year 3 by $\Delta f=0.00072(8)$\,{\cd}. This is a statistically significant change that can also be seen in the $O-C$ diagram. A similar change in period can be seen for the $n=41$ TEO. For the $n=$\,230 TEO, the result is not conclusive because of relatively large errors of the times of maximum light. Since TEOs are expected to have a constant phase and frequency, exactly equal to the integer multiple of the orbital frequency (excluding long-term changes due to the evolution of components and their orbit), we discuss this unexpected result in more detail in Sect.~\ref{sect:discussion}.


\section{The ExtEEV and its TEOs in the ground-based photometry}\label{sect:ground-based}
\subsection{Ground-based photometric data}
\begin{figure*}
   \centering
   \includegraphics[width=0.9\hsize]{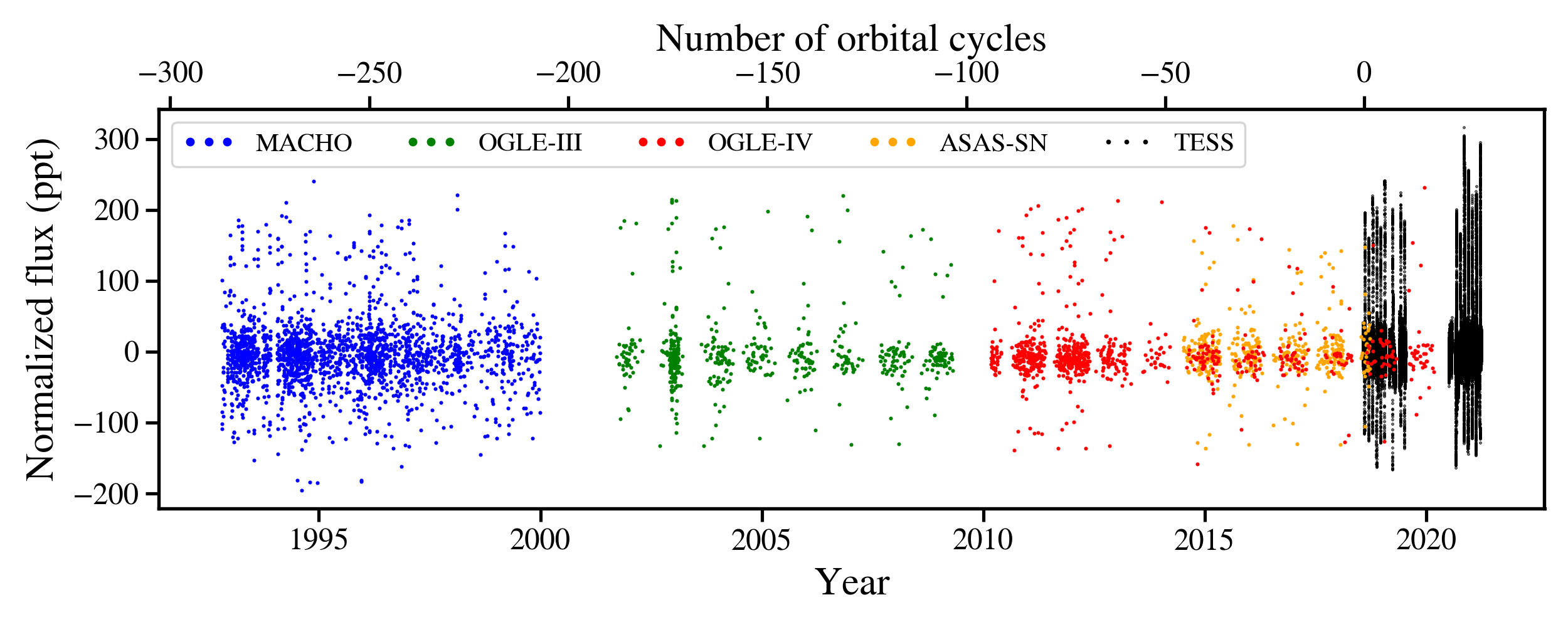}
   \caption{Collection of ground-based photometry and TESS data of the ExtEEV used in the present paper. Thanks to the archival data, it was possible to analyse the behaviour of the heartbeat and the strongest $n=25$ TEO in the ExtEEV for almost 30 years, that is, for over 300 orbital cycles.}
   \label{fig:all-LCs}
\end{figure*}

Having detected changes in amplitudes and periods of TEOs in the ExtEEV, we searched for the ground-based photometry of this star to check if these changes can also be traced on a much longer timescale. Fortunately, because the ExtEEV is located in the LMC, it has a very long record of photometric observations, mainly from the microlensing surveys. Below, we briefly describe these data. Figure \ref{fig:all-LCs} shows the distribution of these observations in time, which together span almost 30 years, without any substantial gaps. 

\subsubsection{MACHO}\label{sect:macho}
Following the idea of \cite{1986ApJ...304....1P}, several extensive surveys were undertaken in the early 1990s aimed at the detection of compact objects in the Galactic halo by means of microlensing. Their detection required observations of as many stars as possible, favouring dense stellar fields such as Magellanic Clouds (MCs) and the Galactic bulge. Regular time-series observations of MCs started in 1992 with the onset of the MACHO survey \citep{1993ASPC...43..291A}. The project brought a wealth of photometric data for millions stars in two bands, $V$ and $R$ \citep{1999PASP..111.1539A}, allowing studies of many types of variable stars. MACHO observations covered the years 1992\,--\,2000. In these observations, the ExtEEV was assigned the MACHO\,80.7443.1718 number and classified as an eclipsing binary with a period of about 32.83~d \citep{2001yCat.2247....0M}. MACHO data of the ExtEEV are available through the project web page\footnote{http://macho.nci.org.au} and CDS. In total, there are 668 and 1487 data points in MACHO $V$ and $R$ passbands, respectively, obtained between October 22, 1992 and January 1, 2000. After some cleaning of outliers, we were left with 650 $V$-filter and 1387 $R$-filter data points. Because no clear difference in the amplitude of the heartbeat between the two passbands was found, the data were combined together. The MACHO light curve of the ExtEEV is shown in Fig.\,\ref{fig:all-LCs},  folded with the orbital period, in Fig.\,\ref{fig:archival-lc+trf}.

\subsubsection{OGLE-III and OGLE-IV}\label{sect:ogle}
OGLE
\citep[][]{uda15} is a long-term large-scale photometric sky survey focused on detecting microlensing events and stellar variability. At present, it monitors the Galactic bulge, Galactic disk, and MCs in Cousins $I$ and Johnson $V$ passbands. The OGLE data used in the present paper were collected with the 1.3-m Warsaw telescope at Las Campanas Observatory, Chile, during the third (OGLE-III, 2001\,--\,2009) and the fourth (OGLE-IV, 2010\,--\,now) phase of the project.

In the present analysis, we use 1451 OGLE-III and OGLE-IV measurements of the ExtEEV obtained in the $I$ passband between 2001 and 2020 (Fig.\,\ref{fig:all-LCs}). The photometry was extracted using custom implementation of the DIA \citep{ala98} method by \citet{woz00}. Typical errors of individual measurements are of the order of 0.005\,mag. OGLE-III and OGLE-IV light curves of the ExtEEV, folded with the orbital period, are shown in Fig.\,\ref{fig:archival-lc+trf}.

\subsubsection{ASAS-SN}\label{sect:asassn}
\begin{figure*}
   \centering
\includegraphics[width=0.85\hsize]{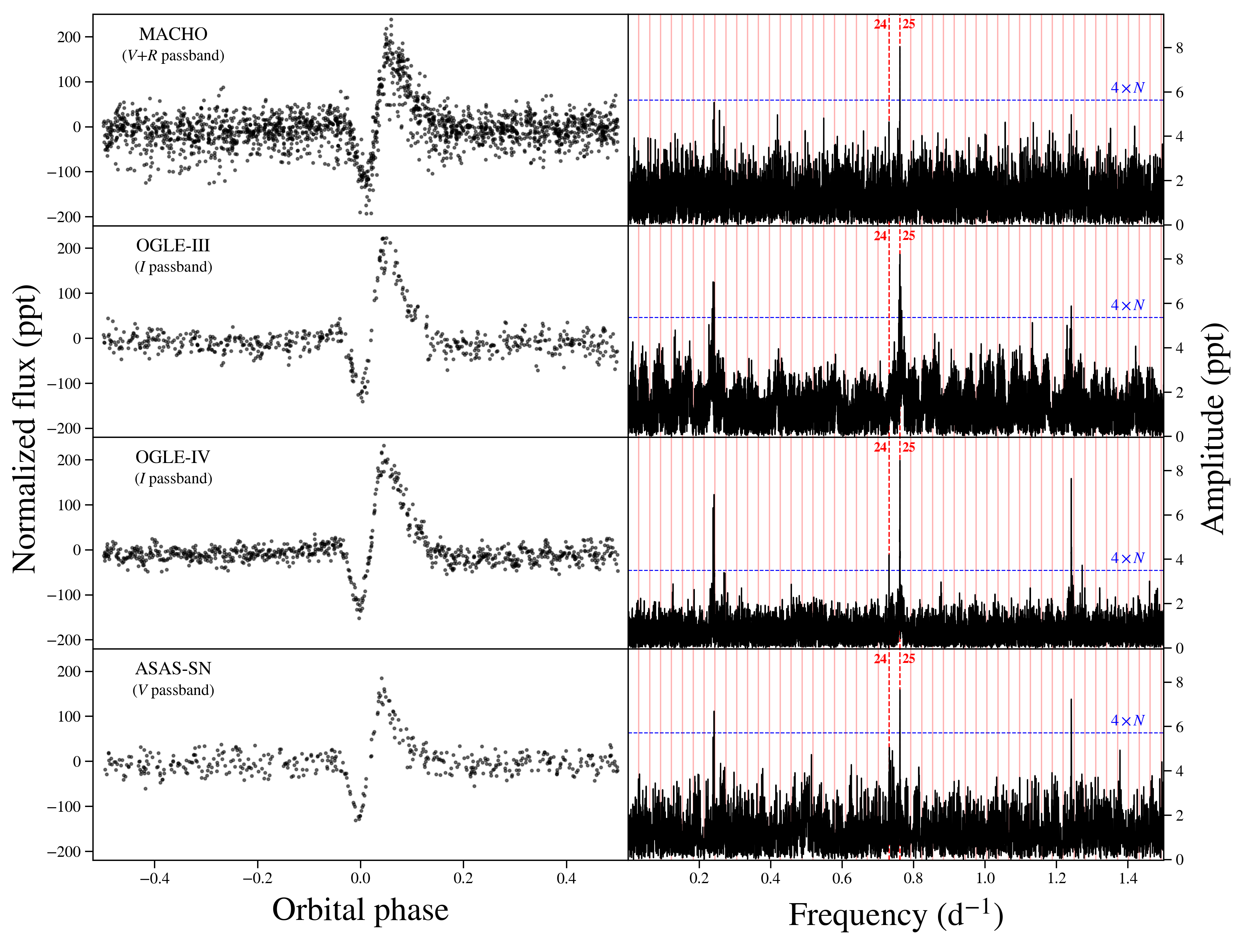}
   \caption{Phased light curves of the ExtEEV from four ground-based projects (left) and their frequency spectra after the subtraction of the heartbeat (right). Blue horizontal lines show the $4\times N$ detection threshold, while the series of vertical solid lines mark the position of consecutive harmonics of the orbital frequency. The pair of thick vertical red lines denote the position of $n=24$ and $n=25$ TEOs. We\ note the presence of daily aliases of the $n=25$ TEO at $\sim$0.24\,{\cd} and $\sim$1.24\,{\cd}. Phase 0.0 corresponds to BJD\,2458373.61518.}
   \label{fig:archival-lc+trf}
\end{figure*}

The ExtEEV was identified as a heartbeat star by Jaya19 during the search for variable stars in the ASAS-SN data. The ASAS-SN project started in 2011 and is primarily aimed at the detection of bright supernovae and other transient phenomena  \citep{2014ApJ...788...48S,2017PASP..129j4502K}. Presently, the survey monitors the entire sky with 24 telescopes down to 18th magnitude in the Sloan $g$ band\footnote{http://www.astronomy.ohio-state.edu/asassn/index.shtml}. We downloaded the ASAS-SN $V$-band light curve of the ExtEEV from the publicly available photometric database of this project\footnote{https://asas-sn.osu.edu/photometry} \citep{2019MNRAS.485..961J}. A few outliers were removed by means of iterative $\sigma$-clipping. Finally, we were left with 438 points in the light curve spread over 4 years, which is shown in Fig.~\ref{fig:all-LCs}. Figure \ref{fig:archival-lc+trf} presents the ASAS-SN light curve phased with the orbital period.

\begin{table}
\caption{Parameters of the $n=24$ and $n=25$ TEOs in ground-based surveys.}
\label{table:archival-pre-whitening}
\centering                         
\begin{tabular}{c l r@{.}l r@{.}l}        
\hline\hline                 
\noalign{\smallskip}
Source & \multicolumn{1}{c}{Frequency} & \multicolumn{2}{c}{Amplitude} & \multicolumn{2}{c}{${\Delta n}^a$} \\
& \multicolumn{1}{c}{(\cd)} & \multicolumn{2}{c}{(ppt)} & \multicolumn{2}{c}{}\\
\noalign{\smallskip}\hline\noalign{\smallskip}
\multicolumn{6}{c}{$n=24$}\\
\noalign{\smallskip}\hline  
\noalign{\smallskip}
MACHO&0.73098(5)&4&8(10)$^\ast$&$-$0&00087(17)\\
OGLE-III& not detected & $<$\,4&7 & \multicolumn{2}{c}{---}\\
OGLE-IV&0.731058(24)&4&7(7)&0&0000(8)\\
ASAS-SN&0.73110(7)&5&4(11)$^\ast$&$+$0&0001(3)\\
\noalign{\smallskip}\hline\noalign{\smallskip}
\multicolumn{6}{c}{$n=25$}\\
\noalign{\smallskip}\hline  
\noalign{\smallskip}
MACHO&0.761495(27)&8&7(10)&$+$0&0009(10)\\
OGLE-III&0.761563(21)&10&5(12)&$+$0&0024(8)\\
OGLE-IV&0.761537(14)&8&3(7)&$+$0&0006(5)\\
ASAS-SN&0.76155(4)&9&1(12)&$-$0&0002(16)\\
\noalign{\smallskip}
\hline                                   
\end{tabular}
\tablefoot{$^a$ $\Delta n=f/f_{\rm orb}-n$, where $f$ is the detected frequency (second column). $^\ast$ Detection below $4\times N$ threshold.}
\end{table}

\subsection{Presence of TEOs}\label{sect:presence-of-teos-archival}
An analysis of individual ground-based light curves revealed the presence of $n=25$ TEO in all four analysed data sets with amplitudes of between 8.3 and 10.5\,ppt (Fig.\,\ref{fig:archival-lc+trf} and Table \ref{table:archival-pre-whitening}). These amplitudes cannot be directly compared because different photometric passbands were used. The only exception are OGLE-III and OGLE-IV data, which were both taken with the Cousins $I$ filter -- a small drop in the amplitude between OGLE-III and OGLE-IV is likely (Table \ref{table:archival-pre-whitening}). It can be seen from Fig.\,\ref{fig:archival-lc+trf} that in addition to $n=25$ TEO, the frequency spectra of MACHO, OGLE-IV, and ASAS-SN data also clearly reveal the presence of the $n=24$ TEO. Although the maxima corresponding to this TEO are below the adopted detection threshold in MACHO and ASAS-SN data, we decided to include them in the variability model. The results of the fitting amplitudes are summarised in Table~\ref{table:archival-pre-whitening}. The detection threshold in the frequency spectra of the ground-based data, ranging between $\sim$3.5\,ppt for OGLE-IV data and $\sim$5.5\,ppt for the other surveys, did not allow us to detect $n=41$ and 230 TEOs if their amplitudes were similar to those in the TESS data (Table \ref{table:separate-pre-whitening}). Apparently, their amplitudes in these observations were below the detection thresholds. The $n=24$ TEO, having an amplitude of $\sim$5\,ppt in the ground-based data (Table \ref{table:archival-pre-whitening}), is only marginally detected in TESS year 1 data and not detected at all in TESS year 3 data (Table~\ref{table:separate-pre-whitening}, Fig.~\ref{fig:tess-2peri-comparison}). This is another indication of the change in amplitudes of TEOs in the ExtEEV.

\subsection{Decrease in the orbital period of the ExtEEV}\label{sect:orbital-period}
Both TESS and ground-based data were subsequently used to verify the stability of the orbital period of the ExtEEV. This was done by means of the $O-C$ diagram using times of minimum light of the heartbeat. Our analysis was based on light curves that were freed from TEOs. In the first step, TESS year 1 data were used to make a template light curve. It was obtained by folding the data with the orbital period, binning in 0.004 to 0.022 phase intervals (shorter in the vicinity of periastron and longer in the flat part of the light curve) and averaging. In total, 80 phase bins were defined. Median values of the phase and flux in phase bins were subsequently interpolated by means of cubic spline functions. A sequence of repeating template light curves was fitted to the most precise set of data, TESS year 1, by applying MCMC methods analogous to those described in Sect.~\ref{sect:stochastic}. As a result, we derived the reference time of minimum light, $T_{\rm ref}=\,\mbox{BJD}\,2458340.66829$, corresponding to the first minimum in the TESS year 1 light curve. As a reference orbital period, we adopted the value derived from all TESS data in Sect.\,\ref{sect:global-pre-whitening}, that is, $P_{\rm ref}=32.83016$\,d. The $O-C$ diagram for the orbital period of the ExtEEV was obtained with these two reference values, that is, assuming $C(E) = T_{\rm ref} + E\times P_{\rm ref}$, where $E$ is the number of orbital cycles that elapsed from $T_{\rm ref}$.

\begin{figure*}
\sidecaption
\includegraphics[width=12cm]{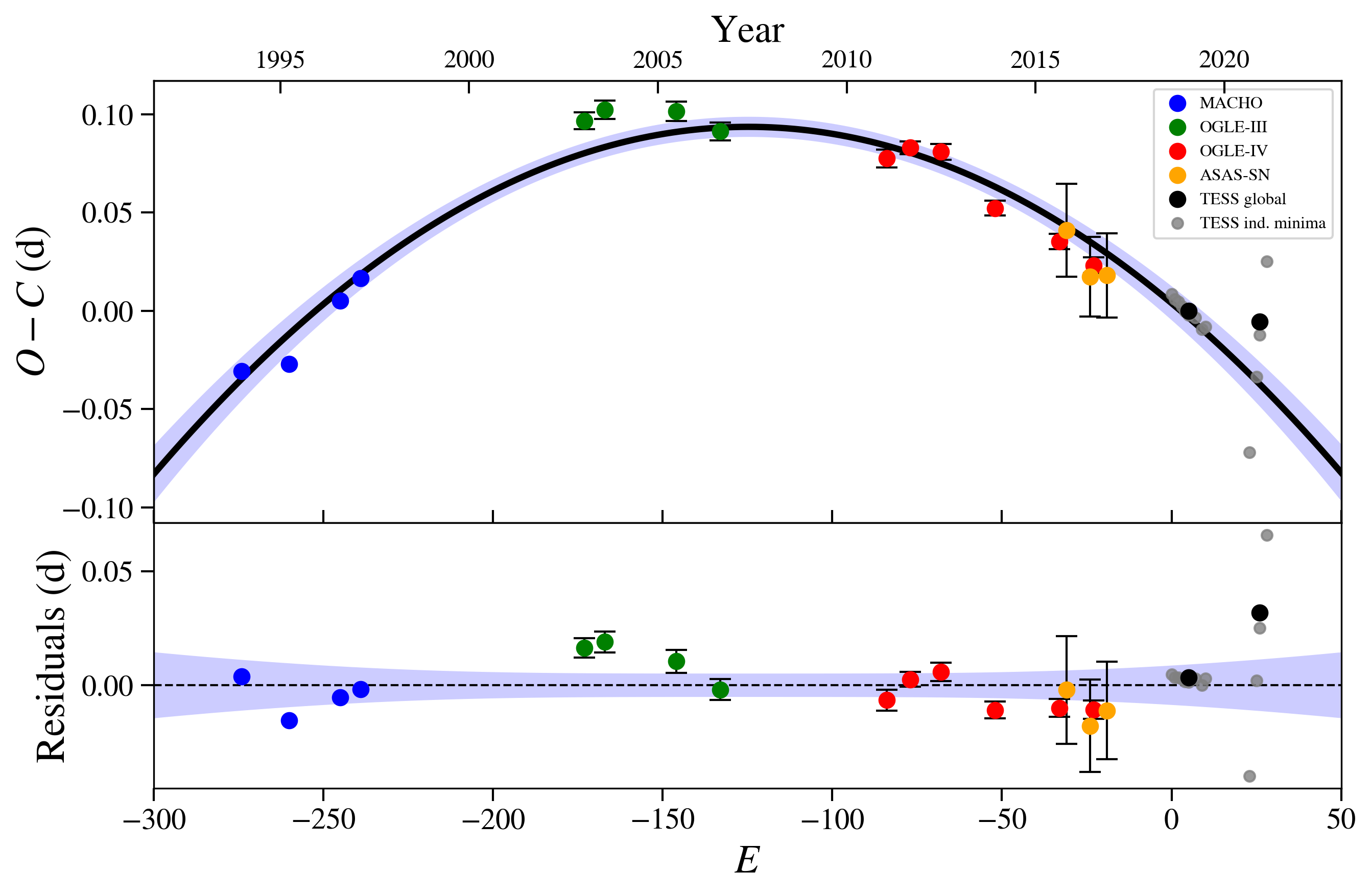}
\caption{$O-C$ diagram for the ExtEEV (top) and residuals from the fitted parabolic model (bottom). The values of $O-C$ for the ground-based surveys and TESS mission are marked with different colours presented in the legend. For TESS data, we provide two types of $O-C$ values. Black points correspond to the global fit to TESS year 1 and 3 data, while grey points represent $O-C$ derived from the observed times of minimum light (bottom part of Table\,\ref{tab:OC_table}). Epoch $E$ stands for the number of orbital cycles that elapsed from the reference time of minimum light, $T_{\rm ref}=\mbox{BJD}\,2458340.66829$. The fitted parabolic model is shown with a solid black line. The blueish shaded areas around the solid and dashed lines denote $1\sigma$ confidence intervals of the best fit.}
    \label{fig:OC_diagram}
\end{figure*}

In the following step, all ground-based data were divided, separately for each survey, into partly overlapping 3 yr-long data samples separated by roughly 1.5\,yr. In total, four MACHO, four OGLE-III, six OGLE-IV, and three ASAS-SN samples were defined. Each sample was first folded with the orbital period. Next, the MCMC methods were used to derive the observed time of minimum light in the folded light curve. This time of minimum light was then transferred to the time of minimum light closest to the average time of a sample by adding or subtracting the integral number of orbital cycles. The derived times of minimum light, $T_{\rm min}=O$, were used to calculate $O-C$ values. They are given in Table~\ref{tab:OC_table} and plotted in Fig.\,\ref{fig:OC_diagram}.
\begin{table}
\centering
\caption{Data presented in the $O-C$ diagram for the orbital period of the ExtEEV (Fig.~\ref{fig:OC_diagram}).}
\label{tab:OC_table}
\begin{tabular}{cr@{.}lrr@{.}l}
\hline
\hline
\noalign{\smallskip}
\multicolumn{1}{c}{Survey} & \multicolumn{2}{c}{$T_{\rm min} -$} & \multicolumn{1}{c}{$E$} & \multicolumn{2}{c}{$O-C$} \\
\multicolumn{1}{c}{or mission}&\multicolumn{2}{c}{BJD 2\,400\,000} & & \multicolumn{2}{c}{(d)}\\
\noalign{\smallskip}
\hline
\noalign{\smallskip}
MACHO&49345&1738&$-$274&$-$0&0307(11)\\
MACHO&49804&7997&$-$260&$-$0&0270(12)\\
MACHO&50297&2842&$-$245&$+$0&0051(13)\\
MACHO&50494&2768&$-$239&$+$0&0167(13)\\
OGLE-III&52661&147&$-$173&$+$0&097(5)\\
OGLE-III&52858&134&$-$167&$+$0&102(5)\\
OGLE-III&53547&567&$-$146&$+$0&102(5)\\
OGLE-III&53974&348&$-$133&$+$0&091(5)\\
OGLE-IV&55583&012&$-$84&$+$0&078(5)\\
OGLE-IV&55812&829&$-$77&$+$0&083(3)\\
OGLE-IV&56108&298&$-$68&$+$0&081(4)\\
OGLE-IV&56633&552&$-$52&$+$0&052(4)\\
OGLE-IV&57257&308&$-$33&$+$0&035(4)\\
ASAS-SN&57322&974&$-$31&$+$0&041(24)\\
ASAS-SN&57552&762&$-$24&$+$0&017(21)\\
OGLE-IV&57585&598&$-$23&$+$0&023(4)\\
ASAS-SN&57716&913&$-$19&$+$0&018(22)\\
TESS year 1& 58504&81909 & 5 & 0&00000(19)\\
TESS year 3& 59194&24690 & 26 & $-$0&00555(15)\\
\noalign{\smallskip}
\hline
\noalign{\smallskip}
TESS&58340&6769&0&$+$0&0086(6)\\
TESS&58373&5043&1&$+$0&0059(6)\\
TESS&58406&3332&2&$+$0&0046(6)\\
TESS&58439&1610&3&$+$0&0022(6)\\
TESS&58471&9887&4&$-$0&0003(6)\\
TESS&58504&8170&5&$-$0&0021(7)\\
TESS&58570&4760&7&$-$0&0034(7)\\
TESS&58636&1302&9&$-$0&0095(6)\\
TESS&58668&9618&10&$-$0&0081(7)\\
TESS&59095&6899&23&$-$0&0720(4)\\
TESS&59161&3888&25&$-$0&0335(4)\\
TESS&59194&2403&26&$-$0&0122(4)\\
TESS&59259&9378&28&$+$0&0251(4)\\
\noalign{\smallskip}
\hline
\end{tabular}
\end{table}

As the TESS data are well sampled, there was no need to phase these sets of data. The times of minimum were therefore derived by using the MCMC analysis of all data surrounding each observed time of minimum. There were nine times of minimum covered by TESS year 1 data and four by TESS year 3 data. These individual times of minimum light for TESS data are provided in the bottom part of Table\,\ref{tab:OC_table} and are marked by `TESS' in the first column. In addition, average times of minimum were derived for the whole TESS year 1 and year 3 data in a similar way as for the ground-based data. These two average times of minimum are given in the middle part of Table\,\ref{tab:OC_table}.

Figure \ref{fig:OC_diagram} shows the resulting $O-C$ diagram. It can be seen that the orbital period, as defined by the times of minimum of the heartbeat, is not constant. In a rough approximation, it can be described by a parabola, which corresponds to a constant rate of period change. Using the first 19 values of $O-C$ from Table\,\ref{tab:OC_table}, we fitted a parabola, $(O-C)(E)=aE^2+bE+c$, by means of the least squares obtaining the coefficient at the quadratic term equal to $a=(-5.8\,\pm\,0.4) \times10^{-6}$\,d per orbital cycle. It can be easily converted to the rate of change in the orbital period, 
\begin{equation}
    \frac{\mbox{d}P_{\rm orb}}{\mbox{d}t}=\dot{P}_{\rm orb}=\frac{2a}{P_{\rm ref}}=(-11.1\,\pm\,0.8)\,\mbox{s}\,\mbox{(yr)}^{-1}.
\end{equation}
The negative value of $\dot{P}_{\rm orb}$ means that the orbital period shortens. We discuss the possible origin of the shortening of the orbital period in Sect.\,\ref{sect:what-causes-changes-of-P}.

As we can seen in Fig.\,\ref{fig:o-c-archival}, the residuals for the times of minimum obtained from TESS year 3 data show large scatter. This can be explained by the influence of the stochastic variability and TEOs. While these effects should average in long time intervals, the individual times of minimum can be affected.

\subsection{Changes in amplitude and frequency of the $n=25$ TEO}\label{sect:changes-of-amplitude-and-frequency-of-the-25-TEO}
\begin{figure*}
   \centering
   \includegraphics[width=0.8\hsize]{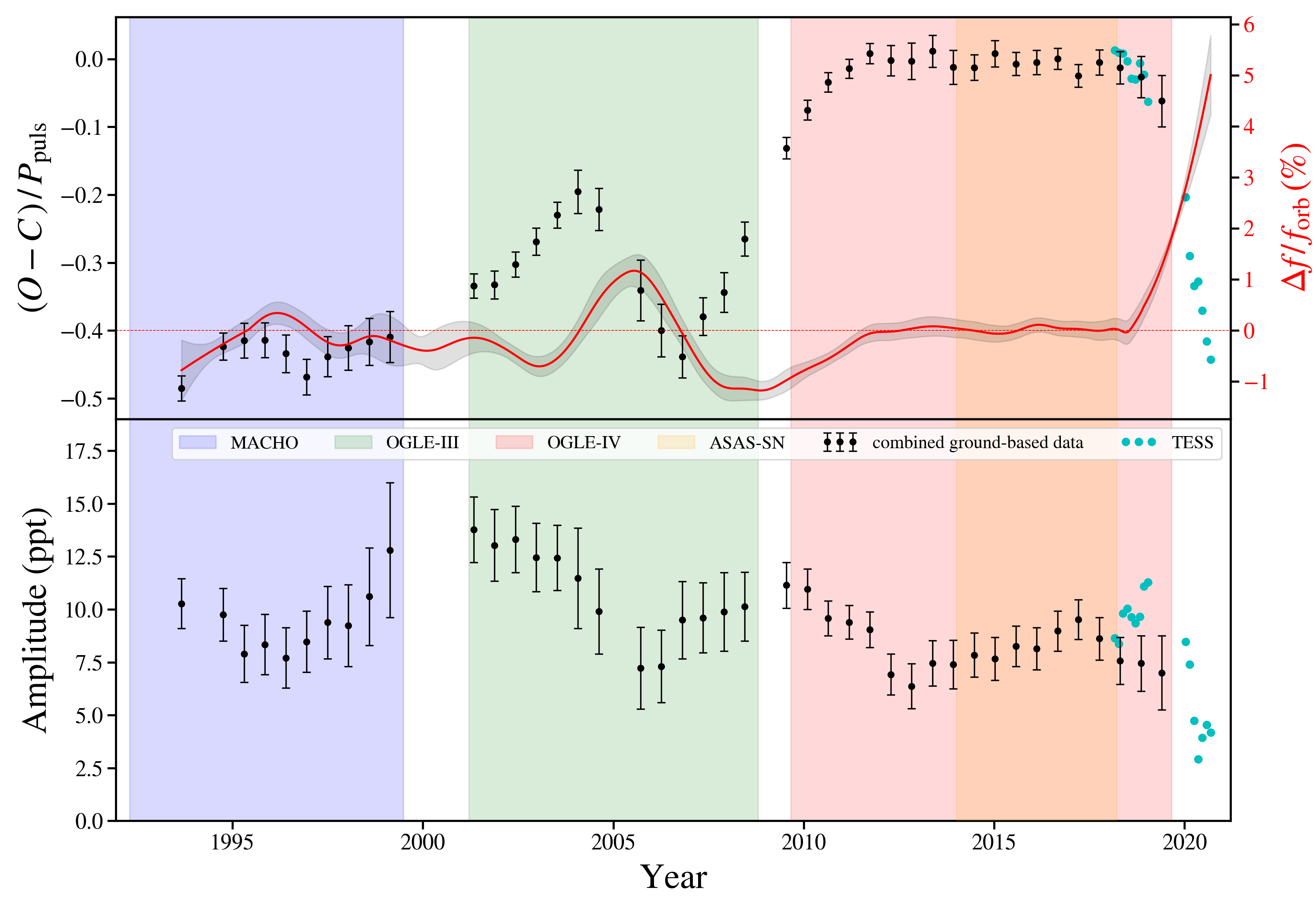}
   \caption{$O-C$ diagram (top) and amplitude changes (bottom) of the $n=25$ TEO observed in the combined ground-based data from the MACHO, OGLE-III, OGLE-IV, and ASAS-SN surveys and the TESS space mission. Vertical stripes mark the time span of the light curves from different ground-based surveys. The solid red curve in the upper panel shows the values of $\Delta f$ defined by Eq.\,(\ref{eq:df}), expressed in terms of the orbital frequency (in per cent). The units of $\Delta f/f_{\rm orb}$ are shown as a right-hand ordinate. The grey shaded area around the curve represents the $\pm1\sigma$ confidence interval derived from the Monte Carlo simulations. More details are given in the main text.}
   \label{fig:o-c-archival}
\end{figure*}

Although the detection threshold in the ground-based data is significantly higher than in the TESS observations, the amplitude of the $n=25$ TEO is large enough to analyse changes in its amplitude and frequency in the same way as we did in Sect.\,\ref{sect:PW-TESS-sliding-window}. In order to minimise the impact of the gaps between different surveys, we combined all ground-based light curves into a single time series. Then, using a sliding window with the width of 1000\,d and a sliding step of 200\,d, we obtained the result which is shown in Fig.\,\ref{fig:o-c-archival}. An order of magnitude longer width of the window than we used in the analysis of the TESS data (Sect.\,\ref{sect:PW-TESS-sliding-window}) means that we can trace changes in the amplitude and frequency of the strongest TEO only on the timescale of years. Faster changes, even if present, are averaged.

Some interesting conclusions can be drawn from Fig.\,\ref{fig:o-c-archival}. First of all, although the amplitude of $n=25$ TEO changes, this particular TEO is present in the ExtEEV for nearly 30 years with an average amplitude of about 10\,ppt (see also Fig.\,\ref{fig:archival-lc+trf}). Second, the most recent $O-C$ values obtained from the OGLE-IV data tend to confirm the trend that can be seen in the TESS data. Third, the change in frequency of the TEO observed in TESS data seems to have a precedent in the past. The OGLE-III data reveal a clear fluctuation in the $O-C$ diagram around 2007, accompanied by a decrease in amplitude from around 13\,ppt  to about 6\,ppt. TESS data show a similar correlation; the pronounced change in frequency of the TEO is followed by a reduction of its amplitude. However, we leave the discussion of this correlation to Sect.~\ref{sect:discussion-changes-of-amplitudes}.

Since both the orbital period (Sect.\,\ref{sect:orbital-period}) and the period of the $n=25$ TEO (this section) change with time, $t$, it is interesting to see how the misalignment of these two values behaves. A good measure of this misalignment is the difference $\Delta f$ defined as follows:
\begin{equation}\label{eq:df}
  \Delta f(t) \equiv f_{\rm puls}(t) - 25f_{\rm orb}(t),
\end{equation}
where $f_{\rm puls}(t)$ and $f_{\rm orb}(t)$ are the pulsation frequency of the $n=25$ TEO and the orbital frequency, respectively. In order to obtain  $f_{\rm puls}(t)$, we first constructed a smooth analytical model of the changes $(O-C)(t)$ seen in Fig.\,\ref{fig:o-c-archival} using the Savitzky-Golay filter \citep{1964AnaCh..36.1627S} of the third order which we then differentiated. The values of $f_{\rm orb}(t)$ were taken assuming a constant value of $\dot{P}_{\rm orb}$ derived in Sect.\,\ref{sect:orbital-period}. The solid red curve in Fig.\,\ref{fig:o-c-archival} shows the derived $\Delta f$ expressed in units of the orbital frequency. As can be seen from this figure, there are two significant excursions from $\Delta f/f_{\rm orb} \sim 0$. The first was of the order of $\pm$1 per cent and took place in the second part of the OGLE-III observations. The other, going up fast to $+5$ per cent is covered by the TESS data and the last part of the OGLE-IV observations.


\section{Discussion}\label{sect:discussion}
\subsection{Occurrence of $n=230$ TEO}\label{sect:why-teo-230}
\begin{figure}
   \centering
   \includegraphics[width=\hsize]{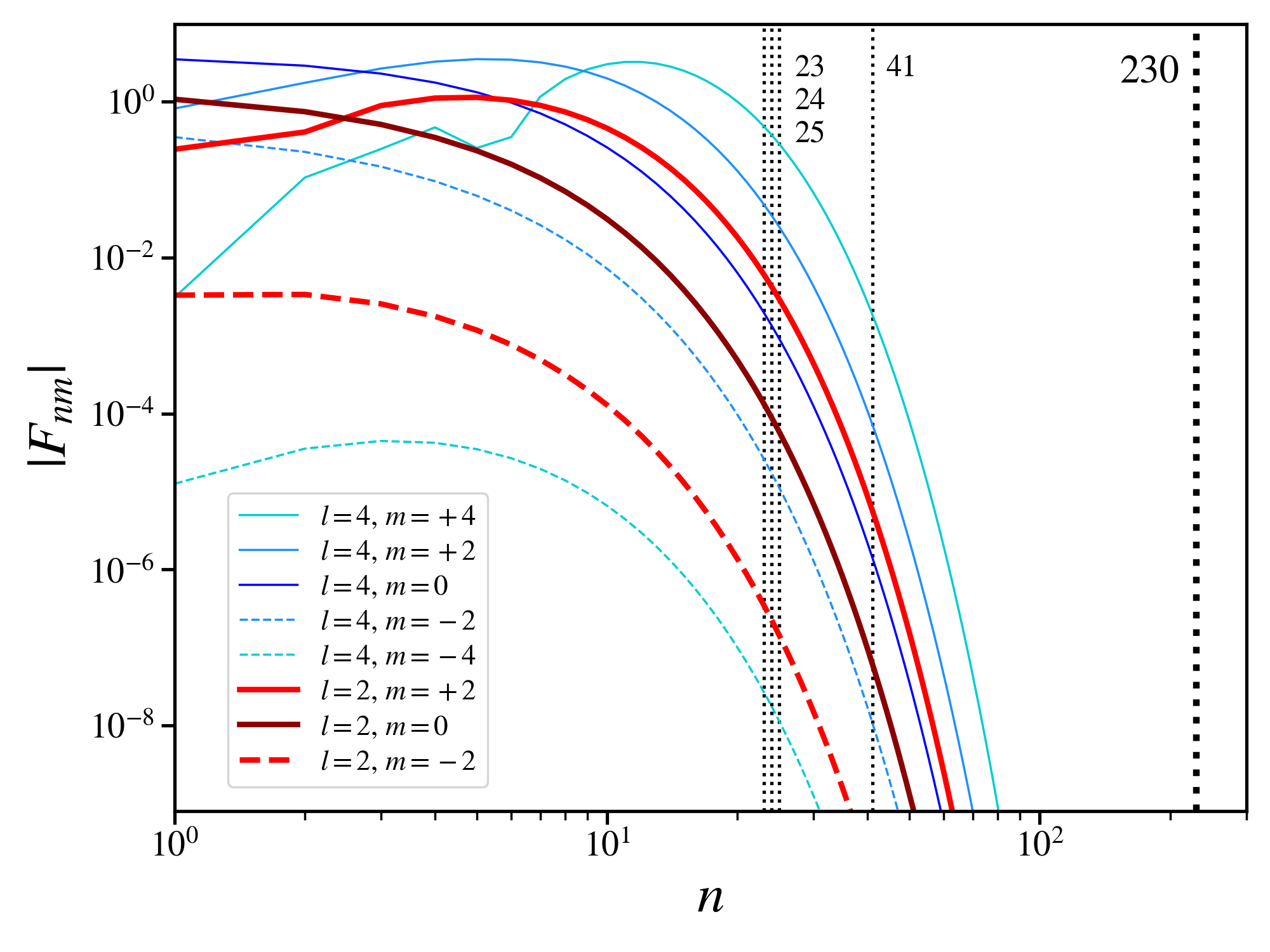}
   \caption{Dependence of Hansen coefficients $F_{nm}$ on the multiple of the orbital frequency, $n$, for TEOs with eight different $(l,m)$ values. The solid and dashed coloured curves show solutions for Eq.~(\ref{eq:hansen}), assuming  $e=$\,0.507 (Jaya21). The vertical dotted lines mark TEOs detected in the ExtEEV. See the main text for more details.}
   \label{fig:hansen-coeff}
\end{figure}

The $n=230$ TEO detected in the ExtEEV is the highest detected multiple of orbital frequency among all TEOs observed in HBSs with masses higher than 2\,M$_\odot$ \citep{2021FrASS...8...67G}. \cite{2017MNRAS.472.1538F} show that the amplitude of TEO (and in consequence the ability of its detection) can be estimated if orbital elements and the seismic model of a star (his eq.~2) are known. One of the crucial factors determining the amplitude of TEO is the equivalent of the so-called Hansen coefficient, $F_{nm}$. It is defined with the following expression (assuming spin-orbit alignment in the system)
\begin{equation}\label{eq:hansen}
    F_{nm}=\frac{1}{\pi}\int\limits_0^\pi\frac{\cos[n(E-e\sin E)-m\upsilon(t)]}{(1-e\cos E)^l}\,{\rm d}E,
\end{equation}
where $n$ is the multiple of orbital frequency, $l$ the degree, $m$ the azimuthal order of spherical harmonic describing the geometry of the TEO, $e$ the eccentricity, $E$ the eccentric anomaly, and $\upsilon(t)$ the true anomaly depending on time $t$. We note that $F_{nm}$ describes temporal coupling between the oscillation mode and characteristic time of the periastron passage. The aforementioned factor expresses an intuitive principle, which says that a mode with the pulsation period close to the characteristic time of periastron passage is most strongly excited. In other words, the higher eccentricity, the higher $n$ is expected to be observed in TEOs. Figure \ref{fig:hansen-coeff} presents the dependence between $F_{nm}$ and $n$ for the orbital eccentricity of the ExtEEV, $e=$\,0.507 (Jaya21), and potentially dominant $l=2$ or $l=4$ TEOs. Of course, one could also consider higher values of $l$ such as $l=5$, 6, or even higher. However, the overall force tidally inducing modes of degree $l$ in a star with radius $R$ scales with $q(R/a)^{(l+1)}$, where $q$ stands for the mass ratio and $a$ denotes the semi-major axis of the orbit. Hence, we do not expect to observe TEOs with high $l$ values.

While $n=$\,23, 24, 25, and 41 TEOs  have a good chance of being excited in the ExtEEV from the point of view of theory, the $n=$\,230 TEO should not be visible at all because it has an extremely small value for its Hansen coefficient\footnote{In principle,  the chance of the excitation of a TEO is more correctly described by the product of $F_{nm}$ and $Q_{nl}$, where $Q_{nl}$ is the so-called overlap integral \citep[cf.~e.g.][his equation 4 and section 9]{2017MNRAS.472.1538F}. The $Q_{nl}$ is proportional to the surface Lagrangian perturbation of the gravitational potential, hence it does not change by more than a few orders of magnitude over a wide range of $n$ for given $l$. Therefore, an extremely small probability of the excitation of the $n=230$ TEO resulting from a very low value of $F_{nm}$ cannot be significantly increased by taking $Q_{nl}$ into account.}. The reason is that for high values of $n$, $F_{nm}$ decays nearly exponentially, which is shown in Fig.\,\ref{fig:hansen-coeff}. Its value for $n=230$ is of the order of $10^{-22}$. The explanation for the occurrence of the $n=230$ TEO, therefore, poses a challenge for theory.

The unusual shape of the light curve of this TEO (Fig.\,\ref{fig:n230-phased-lc}) can possibly be useful in solving the mystery of its origin. For example, it might be related to its high non-adiabaticity. If this is the case, strong non-adiabaticity may enable the TEO to be excited despite its very low value for the $F_{nm}$ coefficient. Theoretical modelling of the pulsation properties of B-type supergiants indicates that their $l=2$ gravity ($g$) modes have frequencies around 1\,--\,2\,{\cd} \citep[e.g.][]{2014IAUS..301..321O}. The $n=230$ TEO with its frequency around 7\,{\cd} occurs in the acoustic frequency range typical for the so-called strange modes \citep[see e.g.][]{1994MNRAS.271...66G,2001RvMA...14..245G,2009CoAst.158..245S}, which are expected to be excited and propagate in the outer envelopes of blue supergiants, where radiation pressure is higher or nearly equal to the gas pressure. These modes are known to be strongly non-adiabatic and their growth rates are comparable to the dynamical timescale. What follows, the observed TEO $n=230$ may be a tidally excited strange mode. We recall that the vast majority of the TEOs observed in known EEVs are $g$-modes with relatively high radial orders \citep{2020ApJ...896..161G}, which is in contrast to the suspected properties of the $n=230$ TEO in the ExtEEV. If the strange character of the mode is confirmed, this particular TEO may provide a unique opportunity to study the dissipation rate of the orbital energy in massive stars through this kind of mode.

\subsection{Variable amplitudes of TEOs}\label{sect:discussion-changes-of-amplitudes}
Almost uninterrupted photometry of the LMC in ground-based surveys and very good quality and cadence of TESS observations allowed us to study the behaviour of amplitudes and periods of TEOs excited in the ExtEEV on a timescale of years and (from TESS data) months. In effect, we provide evidence of variable amplitudes of TEOs. Such variations are known and quite common in massive stars with self-excited pulsations \citep[e.g.][]{1999MNRAS.310..804J,1999NewAR..43..455J,2008A&A...477..907P}. They are possibly caused by non-linear mode coupling, where the energy of one mode is transferred to the other coupled mode(s) \citep{1982AcA....32..147D,1985AcA....35....5D,1988AcA....38...61D}. In the case of non-linear resonance mode coupling, TEOs can have frequencies that are not close to a harmonic of the orbital frequency, but instead they sum up to a harmonic \citep[e.g.][]{2012MNRAS.421..983B,2020ApJ...896..161G}. For the ExtEEV, it is hard to tell if there are any non-harmonic low-amplitude modes excited by TEOs because of the presence of the stochastic variability. It raises the signal at low frequencies and makes the detection of potential daughter modes\footnote{In the case of a three-mode resonance, frequencies of two daughter modes sum up to the frequency of the parent or mother mode.} merely possible (cf.~Fig.\,\ref{fig:tess-peri-all}, panels b and c). The $n=23$, 24, 25, and 41 TEOs would have their daughter modes in the region occupied by $g$-modes. Their growth rates are significantly longer than a month, which is the timescale of amplitude changes detected in TESS data (Fig.\,\ref{fig:teo-ampl-oc}). However, a system of three resonantly coupled modes even with very different linear growth rates can behave in many ways, including exponential growth, an equilibrium solution, chaos, and limit cycles \citep{1985AcA....35..229M}. The latter ones, despite their potentially long periods, may occasionally exhibit a rapid decrease or increase in amplitude on a timescale comparable to that seen in ExtEEV. The more generalised amplitude equations for several TEOs interacting with each other (including non-linear tidal effects) and for any number of daughter modes are, however, more complicated and given, for example,~by \cite{2012ApJ...751..136W}. Periodic amplitude fluctuations are still among the possible solutions that may explain the behaviour of TEOs observed in ExtEEV.

We would also like to consider another explanation. Figures \ref{fig:teo-ampl-oc} and \ref{fig:o-c-archival} show that at least $n=25$ and 41 TEOs exhibit significant changes in their frequencies. It is particularly pronounced for the dominant $n=25$ TEO (upper panel of Fig.\,\ref{fig:o-c-archival}). It seems from this figure that at least for the TESS data, there is a correlation between $\Delta f/f_{\rm orb}$ and amplitude: the higher the $\Delta f/f_{\rm orb}$, the lower the amplitude. It is possible that similar correlations also exist for the remaining TEOs observed in the ExtEEV, although this cannot be verified with the available data. Therefore, it seems reasonable to conclude that small variations in frequencies of TEOs cause changes in their amplitudes. This is in agreement with the theory \citep[e.g.][]{2017MNRAS.472.1538F} that predicts that amplitudes of TEOs are sensitive to the detuning factor, which depends on the difference between the eigenfrequency of a mode and the nearest harmonic of the orbital frequency\footnote{We would like to emphasise that the aforementioned difference is not identical to the `observational' $\Delta f$ introduced in Eq.~(\ref{eq:df}). Due to non-linear effects, a TEO never has a frequency strictly equal to $nf_{\rm orb}$. Therefore, it is difficult to convert $\Delta f$ determined by us to the detuning factor, which relates to the spectrum of the normal modes. However, these two differences are related in the sense that a change in one of them entails a change in the other.}. Even small changes in the mode frequency may cause large changes in the detuning factor and TEO's amplitude. In this scenario, we would observe the effect of detuning between the eigenfrequency of the stellar mode and the nearest harmonic of the orbital frequency. The eigenmode would be tidally excited as long as its frequency is very close to the orbital harmonic. When they are moving apart, the efficiency of tidal forcing decreases rapidly and the mode decouples from the influence of a given orbital harmonic. The amplitude of such a decoupled mode would decrease on a timescale dictated by its growth rate. A reverse situation may occur when the frequency of an eigenmode approaches the orbital harmonic --- an increase in amplitude of a TEO should be observed. 

In blue supergiants, non-adiabatic effects in pulsations are important. Therefore, we may suspect that the observed TEOs have relatively short growth rates, compatible with the timescale of their amplitude changes reported in the present paper. Even small mass losses or changes in the extended atmosphere of a blue supergiant may lead to slight changes in pulsation frequencies and, in consequence, changes in the detuning parameter and amplitudes of the TEOs. Next, blue supergiants have a thin convective layer approximately halfway between the core and the stellar surface, overlying the hydrogen burning shell \citep{2009A&A...498..273G}. Turbulent and large-scale mass motions in this supra H-shell or intermediate convection zone may also contribute to the small changes in frequencies of eigenmodes. While a detailed study of the related timescales is beyond the scope of this paper, it is possible that small excursions of frequencies of TEOs from the exact resonance with the orbital harmonic occur naturally in supergiants. At least in the primary of the ExtEEV, we clearly see such changes, while in the A-type binary HD\,181850, \cite{2020ApJ...896..161G} found no evidence of variable amplitudes of TEOs over four years of Kepler observations. Whether there are differences in the long-term stability of amplitudes and periods of TEOs in blue supergiants and early-type dwarfs has yet to be established with many more examples of EEVs with TEOs.

\subsection{Changes in the orbital period}\label{sect:what-causes-changes-of-P}
We found in Sect.\,\ref{sect:orbital-period} that the orbital period of the ExtEEV, as traced by the times of minimum of the heartbeat, shortens with a mean rate of about 11\,s\,(yr)$^{-1}$. Searching for the possible explanation of this change, we considered five different phenomena: emission of gravitational waves, apsidal motion, light-travel time effect, tidal dissipation of orbital energy, and mass transfer in the system.

In order to estimate the effect of emission of gravitational waves on the observed $\dot{P}_{\rm orb}$, we used eq.~(5.6) from the work of \cite{1964PhRv..136.1224P}. By adopting physical parameters of the ExtEEV obtained by Jaya21, we obtained $\dot{P}_{\rm orb}\approx -$1.5$\,\times\,10^{-6}$\,s\,(yr)$^{-1}$. This value is seven orders of magnitude lower than the observed one. We conclude that the impact of the emission of gravitational waves on the changes in the orbital period is negligible in the ExtEEV and cannot explain the observed value.

Apsidal motion can result in apparent changes in the orbital period measured with the times of minimum light of eclipses provided that the data cover a significant part of the period of apsidal motion. The ExtEEV is a non-eclipsing system, but the apsidal motion can be traced with the heartbeat because its shape depends on $\omega$, the argument of periastron. In general, apsidal motion leads to the advance of $\omega$. Around the measured value of $\omega\approx 302\degr$ (Jaya21) this should lead, with time, to a decrease in the depth of the minimum and increase in the height of the following maximum of the heartbeat. We found no evidence of a change in the shape of the heartbeat for about 19 years of the $I$-filter OGLE observations. This means that during this time interval, $\omega$ did not change more than $\sim$2$\degr$. This gives an upper limit for d$\omega$/d$t=\dot{\omega}\lesssim$\,0.1$\degr$\,(yr)$^{-1}$. We also estimated a theoretical value of $\dot{\omega}$ following \cite{2020A&A...642A.221R}. Using physical parameters of the ExtEEV from Jaya21 and assuming internal structure constants $k_{2,1}=0.00125$ and $k_{2,2}=0.01$ from \cite{2019A&A...628A..29C}, we obtained $\dot{\omega}\approx 0.082\degr$\,(yr)$^{-1}$, in full consistency with the upper limit obtained from the observations. The resulting period of apsidal motion $U=360\degr/\dot{\omega}\approx 4400$\,yr is more than two orders of magnitude longer than the time span of the ground-based observations of the ExtEEV. This means that the latter covers only a very small part of the period of apsidal motion, and that apsidal motion cannot explain changes seen in Fig.\,\ref{fig:OC_diagram} either.

Another phenomenon that can explain the shortening of the orbital period is the presence of an additional body (or bodies) that would cause a light-travel time effect \citep[hereafter LTTE,][]{1952ApJ...116..211I}. We do not know what the orbital parameters are of this potential tertiary component. The only limitation we have is related to its orbital period, which must be longer than $\sim$30 years. We can therefore estimate the minimum mass of the tertiary assuming that the inclination of its orbit is 90{\degr} and considering different values of eccentricity. By fitting the LTTE curves described within the formalism provided by \cite{1952ApJ...116..211I} to the data presented in Fig.~\ref{fig:OC_diagram}, we obtained a low limit on the tertiary's mass, which is equal to about 18\,M$_\odot$. This is a value comparable to the mass of the secondary estimated by Jaya21. Since the secondary's lines were not detected in the spectra of ExtEEV, the potential tertiary's lines might have also remained undetected. This means that ExtEEV can be a multiple hierarchical system and the presence of a tertiary can explain the change in the orbital period seen in Fig.~\ref{fig:OC_diagram}. 

The remaining two mechanisms considered here, tidal dissipation of orbital energy and mass transfer in the system, are also able to explain the observed change in the orbital period, but the question arises whether they are at work. Let us start with tides. They can transform orbital angular momentum into the spin angular momenta of components and vice versa (spin-orbit coupling). Tides have also a non-conservative nature because they can transform orbital energy into thermal energy of stellar interiors. Assuming that tidal dissipation of the orbital energy is the only source of a decreasing orbital period, the observed value of $\dot{P}_{\rm orb}$ can be converted into effective tidal quality factor $Q$ \citep{1966Icar....5..375G,2017MNRAS.472L..25F}, which for the ExtEEV amounts to about 1.2\,$\times\,10^{2}$. However, as shown by \cite{1999A&A...350..129W} and \cite{2017MNRAS.472L..25F}, $Q$ may exhibit significant changes on relatively short timescales, depending on the resonance locking conditions and the number of excited TEOs. A simple estimation of the orbital decay timescale defined as $\tau_{\rm d}\equiv a/\dot{a}$, where $a$ is the semi-major axis of the system's orbit, returns $\tau_{\rm d}\approx 4\,\times\,10^5$\,yr. This means that if a tidal dissipation scenario is effective, the orbit of the ExtEEV evolves on a timescale comparable to the nuclear timescale of the primary.

As we claimed above, mass transfer in the system can also explain the observed value of $\dot{P}_{\rm orb}$. From the observational point of view, the presence of mass transfer in the ExtEEV is justified by evidence of the presence of a disk that changes its size with the orbital phase (Jaya21). Neither Jaya19 nor Jaya21 were able to reproduce the amplitude of the heartbeat in the ExtEEV using the \texttt{PHOEBE\,2} code \citep{2016ApJS..227...29P,2018ApJS..237...26H,2020ApJS..247...63J,2020ApJS..250...34C}. We confirm this result. Assuming parameters of the system derived by Jaya21 with their estimate of the primary's radius equal to $\sim$24\,R$_\odot$, we obtained the peak-to-peak amplitude of the heartbeat equal to about 10\,ppt. Next, we changed the primary's radius to 37.5\,R$_\odot$, which corresponds to a configuration in which the primary nearly fills its Roche lobe at periastron. It allowed us to increase the resulting amplitude of the heartbeat to about 50\,ppt. This exercise shows two things. First, the inability of reproducing the heartbeat with standard models means that some additional contribution to the total flux, presumably from the disk, is required. Next, if the true primary's radius is larger than 24\,R$_\odot$, the conditions for mass transfer close to periastron become even better. A mass transfer in the system is therefore very likely, especially because the supergiant has an extended envelope. Assuming conservative mass transfer from the primary to the secondary and keeping the total orbital angular momentum constant, we found that the required mass transfer rate that would reproduce the observed $\dot{P}_{\rm orb}$ in the ExtEEV amounts to about 4\,$\times\, 10^{-5}$\,M$_\odot$\,(yr)$^{-1}$. 

The loss of mass and total orbital angular momentum by the tidally enhanced stellar wind may also contribute to $\dot{P}_{\rm orb}$ observed in the ExtEEV. This mass loss should be phase-dependent and especially enhanced during periastron passages \citep[e.g.][]{1988MNRAS.231..823T,2006epbm.book.....E}. If stellar wind is strong enough near periastron, it can delay the circularisation of a system or even increase eccentricity \citep{2000A&A...357..557S}. We estimated the mass-loss rate assuming that only tidally enhanced stellar wind of the primary is responsible for the decreasing orbital period. We treated the entire phenomenon as perfectly non-conservative, that is, for each orbital period, a portion of the matter was ejected from the primary outside the system with the angular momentum estimated from the primary's velocity at periastron. We found that the mass loss rate of about $10^{-4}$\,M$_\odot$\,(yr)$^{-1}$ is necessary to explain the observed $\dot{P}_{\rm orb}$. This is a large value, but TEOs present in the ExtEEV can intensify stellar wind, as indicated, for example,~by \cite{2007AIPC..948..345T} and \cite{2015A&A...581A..75K} for pulsating massive stars. Moreover, \cite{2017MNRAS.471.3245Y} show that strange mode instabilities in massive OB-type stars have the potential to induce stellar winds with mass-loss rates reaching $10^{-4}$\,M$_\odot$\,(yr)$^{-1}$. Regardless of the mechanism which drives stellar wind in the ExtEEV, the ejected material can form a circumbinary disk, which also can affect the orbit of the ExtEEV, including the evolution of its eccentricity \citep{1991ApJ...370L..35A}.
    
The above described effects of mass loss, mass transfer, and tidal dissipation in the ExtEEV can explain the observed value of $\dot{P}_{\rm orb}$. The effects are difficult to separate, but most likely they all contribute to the observed shortening of the orbital period. Therefore, further high-resolution spectroscopic studies are needed to distinguish them.

\section{Summary and conclusions}
The present study focuses on the detection and characterisation of TEOs in the extreme-amplitude EEV,  MACHO\,70.7443.1718, a very massive (35 + 16\,M$_\odot$) system in the LMC in a highly eccentric ($e\approx 0.5)$ orbit (Jaya21). In particular, we concentrated on the temporal behaviour of amplitudes and periods of the TEOs. Our most important findings can be summarised as follows:
\begin{enumerate}
 \item We confirm the presence of $n=23$, 25, and 41 TEOs found by Jaya19 and Jaya21 (Sect.~\ref{sect:global-pre-whitening}). We also found two additional TEOs at $n=24$ and 230 (Sects.~\ref{sect:global-pre-whitening} and \ref{sect:TESS-separate-prewhitening}). 
 \item The $n=230$ TEO is the highest harmonic of all TEOs detected thus far in massive ($\gtrsim 2\,M_\odot$) stars \citep{2021FrASS...8...67G}. The TEO is also unique in other respects. First,  such a high-$n$ TEO is not predicted to be excited in a system with $e\approx 0.5$, according to the present theory (Sect.~\ref{sect:why-teo-230}). Second, it has a strongly non-sinusoidal light curve (Fig.~\ref{fig:n230-phased-lc}). We speculate that with a frequency of $\sim$7\,\cd, falling into the range of acoustic modes, the TEO can be a strongly non-adiabatic $p$-mode or even a strange mode. This makes it extremely interesting as the majority of known TEOs in main-sequence stars are $g$-modes.
 \item We show that amplitudes of TEOs in the ExtEEV significantly change on a timescale of months and years (Sect.~\ref{sect:changing-a-and-f}). This is the first piece of evidence of rapidly changing amplitudes of TEOs. The changes are particularly well pronounced for the dominant $n=25$ TEO, which reduced its amplitude twofold during only four orbital cycles (Fig.~\ref{fig:teo-ampl-oc}). Changes in amplitudes were also detected for other TEOs (Sect.~\ref{sect:presence-of-teos-archival}).
 \item In addition to changing amplitudes, we show that $n=25$ and 41 TEOs change their frequencies (Sect.~\ref{sect:PW-TESS-sliding-window}). It seems, at least for the $n=25$ TEO, that changes in amplitudes and frequencies are related (Sects.~\ref{sect:changes-of-amplitude-and-frequency-of-the-25-TEO} and \ref{sect:discussion-changes-of-amplitudes}). We interpret this behaviour as the effect of the non-linear mode coupling or detuning between the frequencies of oscillation modes of the primary component and harmonics of the orbital frequency. We also discuss why this behaviour may occur in blue supergiants (Sect.~\ref{sect:discussion-changes-of-amplitudes}). 
 \item The orbital period of the ExtEEV decreased during the past $\sim$30 years at a rate of about 11\,s\,(yr)$^{-1}$ (Sect.~\ref{sect:orbital-period}). We indicate that a plausible explanation of these changes involves the presence of a tertiary on a long (orbital period longer than 30~yr) orbit. If this is the case, the ExtEEV forms a hierarchical triple system. Mass transfer between the components and mass loss from the system due to stellar wind(s) can also explain the change in the orbital period. If these processes are efficient, mass transfer or mass loss rates of the order of $10^{-5}$\,--\,$10^{-4}$\,M$_\odot$\,(yr)$^{-1}$ are required (Sect.~\ref{sect:what-causes-changes-of-P}). Tidal dissipation of the orbital energy with high-amplitude TEOs may also contribute to the shortening of the orbital period.
\end{enumerate}

Amplitude and frequency changes can be quite common among EEVs with TEOs. At present, we have a detection of relatively fast changes in amplitudes and frequencies of TEOs only in one star, the ExtEEV. Much slower changes, of the order of 2\,--\,3\% during four years of Kepler observations, were found by \cite{2014MNRAS.440.3036O} in KOI-54, the archetype of EEVs. Next, \cite{2020ApJ...896..161G} found no evidence of changing amplitudes and frequencies in a system composed of two A-type components. The question is whether the circumstances for amplitude changes in A-type stars are different for B-type supergiants or maybe whether timescales of these changes for A-type stars are much longer. Some general conclusions as to the causes and consequences of amplitude and frequency changes can be formulated only when their stability in a large sample of EEVs with TEOs will be studied. Our results, therefore, encourage observational studies of TEOs in EEVs, and pose a challenge to the theory. Thus, we would like to conclude our paper with some important problems that stem from our findings.
\begin{enumerate}
    \item The excitation mechanism for the $n=230$ TEO is not obvious and requires theoretical explanation. In Sect.~\ref{sect:why-teo-230} we argued that the value of the $F_{nm}$ coefficient does not allow for an excitation of such a high-$n$ TEO in the ExtEEV system. In general, the dependence of the TEO's amplitude on the $F_{nm}$ coefficient should effectively prevent tidal excitation of $p$ modes, except for orbits with extremely high eccentricities. If an effective mechanism for excitation of high-$n$ TEOs is found, we can expect TEOs to occur in a much wider range of frequencies than previously thought, including the ranges of $p$- and strange modes.
    \item The question of how does the presence of strongly non-adiabatic $p$-mode or strange mode TEOs influence dynamical evolution of massive binary systems remains an open question. Significantly non-adiabatic pulsations have very short growth rates, which can be comparable to the orbital period. This means that in a situation that favours resonance, the amplitude of such a TEO can notably increase in less than a few orbital cycles. This, in turn, may result in a rapid drop in the tidal quality factor and quick shrinkage of the orbit.
    \item It is not clear why frequencies of TEOs in the ExtEEV (and generally in blue supergiants) change on a timescale of months. We showed that at least $n=25$ and 41 TEOs change their frequencies significantly, but there are indications that the other TEOs in the ExtEEV may behave in a similar way.
    \item In massive stars, TEOs can affect the evolution of the orbit in two ways: firstly, by a direct dissipation of the total orbital energy in stellar interiors and, secondly, by taking out the orbital and rotational angular momenta because of the stellar wind driven or enhanced by pulsations. The coupling between the intensity of stellar wind and the high-amplitude TEOs, which are mainly propagating in the outer envelope of the primary of the ExtEEV, needs further investigation.
\end{enumerate}

The ExtEEV is an excellent target to address the questions raised above. The role of TEOs in the dynamical evolution of massive binary systems is still a matter of large uncertainties. However, TEOs represent a potentially effective and still unexplored channel for draining the orbital energy from the system on timescales comparable to the nuclear timescales of the components. In some circumstances, this can possibly lead to the formation of a contact binary or even tidal disruption and coalescence of massive stars.


\begin{acknowledgements}
We would like to thank Professor Michael D.~Albrow for patient and fruitful discussion on the application of \texttt{pyDIA} code to the TESS FFIs. PKS,  AP, MR, and MW have been supported by the Polish National Science Center grants no.~2019/35/N/ST9/03805, 2016/21/B/ST9/01126, 2016/22/A/ST9/00009, and 2018/31/B/ST9/00334, respectively. The authors made use of Strasbourg astronomical Data Center (CDS) portal and Barbara A.~Mikulski Archive for Space Telescopes (MAST) portal. This paper includes data collected by the TESS mission, which are publicly available from the MAST. This research has made use of "Aladin sky atlas" developed at CDS, Strasbourg Observatory, France. This paper utilises public domain data originally obtained by the MACHO Project, whose work was performed under the joint auspices of the U.S.~Department of Energy, National Nuclear Security Administration by the University of California, Lawrence Livermore National Laboratory under contract No.~W-7405-Eng-48, the National Science Foundation through the Center for Particle Astrophysics of the University of California under cooperative agreement AST-8809616, and the Mount Stromlo and Siding Spring Observatory, part of the Australian National University. This research has made use of the VizieR catalogue access tool, CDS, Strasbourg, France. The original description of the VizieR service was published by \cite{2000A&AS..143...23O}.
\end{acknowledgements}

%
%

\bibliographystyle{aa}                                                  
\bibliography{PKS_ExtEEV}

\end{document}